\documentclass[revtex4]{emulateapj}
\usepackage[T1]{fontenc}
\usepackage[latin9]{inputenc}
\usepackage{amsmath} 
\usepackage{amsfonts}
\usepackage{amssymb}
\usepackage{graphicx}
\usepackage{lmodern}
\usepackage[pdftitle={SMBH Mass Assembly Via Accretion and Mergers}, pdfauthor={Kulier et al.}]{hyperref}
\newcommand{\der}{\mathrm{d}}

\shorttitle{SMBH Mass Assembly Via Accretion and Mergers}
\shortauthors{Kulier et al.}

\begin{document}

\title{Understanding black hole mass assembly via accretion and mergers at late times in
cosmological simulations}

\author{Andrea Kulier\altaffilmark{1}}
\email{akulier@princeton.edu}
\author{Jeremiah P. Ostriker\altaffilmark{1,2}}
\author{Priyamvada Natarajan\altaffilmark{3}}
\author{Claire N. Lackner\altaffilmark{1,4}}
\author{Renyue Cen\altaffilmark{1}}

\altaffiltext{1}{Department of Astrophysical Sciences, Princeton University, Princeton, NJ 08544, USA}
\altaffiltext{2}{Department of Astronomy, Columbia University, NYC, NY 10027, USA}
\altaffiltext{3}{Department of Astronomy, 260 Whitney Avenue, Yale University, New Haven, CT 06511, USA}
\altaffiltext{4}{Kavli Institute for the Physics and Mathematics of the Universe (WPI), 
Todai Institutes for Advanced Study, the University of Tokyo, Kashiwa, Japan}

\begin{abstract}
Accretion is thought to primarily contribute to the mass accumulation 
history of supermassive black holes throughout cosmic time. While this
may be true at high redshifts, at lower redshifts and for the most
massive black holes mergers themselves might add significantly to the mass budget. 
We explore this in two disparate environments --- a massive cluster and a void region. 
We evolve SMBHs from $4 > z > 0$ using merger trees 
derived from hydrodynamical cosmological simulations of these two regions,
scaled to the observed value of the stellar mass fraction to account for overcooling.
Mass gains from gas accretion proportional to bulge growth and BH-BH mergers are tracked,
as are black holes that remain ``orbiting'' due to insufficient dynamical 
friction in a merger remnant, as well as those that are ejected due to
gravitational recoil. We find that gas accretion remains the dominant source of mass accumulation
in almost all SMBHs; mergers contribute $2.5\pm0.1\%$ for all SMBHs in the
cluster and $1.0\pm0.1\%$ in the void since $z = 4$. However, mergers are significant
for massive SMBHs. The fraction of mass accumulated from mergers for central black holes 
generally increases for larger values of 
the host bulge mass: in the void, the fraction is $2\%$ at $M_{*, \mathrm{bul}} = 10^{10} M_{\odot}$,
increasing to $4\%$ at $M_{*, \mathrm{bul}} \gtrsim 10^{11} M_{\odot}$, and
in the cluster it is $4\%$ at $M_{*, \mathrm{bul}} = 10^{10} M_{\odot}$ and $23\%$ at $10^{12} M_{\odot}$.
We also find that the total mass in orbiting SMBHs is negligible in the void, but significant in the cluster,
in which a potentially detectable $40\%$ of SMBHs and $\approx 8\%$ of the total SMBH 
mass\footnote{where the total includes central, orbiting,
and ejected SMBHs} is found orbiting at $z = 0$. 
The existence of orbiting and ejected SMBHs 
requires modification of the Soltan argument. We estimate this correction to 
the integrated accreted mass density of SMBHs to be in the range $6-21\%$, 
with a mean value of $11\pm3\%$. 
Quantifying the growth due to mergers at these late times, we calculate the total 
energy output and strain from gravitational waves emitted by merging SMBHs, and
obtain a signal potentially detectable by pulsar timing arrays.
\end{abstract}

\keywords{black hole physics --- galaxies: nuclei --- quasars: general}

\section{Introduction}

Observations strongly suggest that supermassive black holes are harbored in the centers of almost all massive
galaxies \citep{kr1995}.  The masses of these central black holes are observed to correlate with
multiple properties of their host galaxies. The most well-known and studied of these relations
is the $M-\sigma$ relation between the mass of the black hole and the velocity dispersion of
its host galaxy spheroid \citep{g2000, fr2000, tetal2002, getal2009}. Relations between the
black hole mass and the stellar mass and luminosity of the bulge have also been derived from
observations  \citep{mh2003, hr2004, bei2012, mcc2012}. It has also been claimed that the mass of the central
black hole is correlated with galactic properties on scales larger than the bulge,
such as the total stellar mass \citep{ci2011, b2011, bei2012} and luminosity \citep{lasker2014}
of the host galaxy, and the mass of the entire host halo (\citealt{ferr2002, vng2011}; see however \citealt{kb2011a, kb2011b}).
There is also evidence of increased scatter in some of these relations for low mass galaxies
\citep{gree2010, vng2011, mcc2012}. The existence of these correlations suggests interplay
between black hole growth and star formation activity in galactic nuclei.

There appears to be a connection between AGN activity and star formation on
galactic scales larger than the nucleus as well. The
total AGN activity follows roughly the same trend as a function of cosmic redshift as the global
star formation rate \citep{heckman2004, merloni2004, brusa2009, shankar2009}. A correlation has
also been found between the luminosity of individual AGN and the star formation luminosity of
their host galaxies \citep{netz2009, chen2013}. Some
have found that this correlation weakens or disappears for low-luminosity AGN
and at $z \gtrsim 1$ \citep{rosario2013}, but others claim that
these observations can be explained if mean AGN activity
follows host galaxy SFR on timescales $\gtrsim 100$ Myr
but is highly variable on shorter timescales \citep{hickox2014}. 
It is becoming clear at high redshifts that the global 
star formation rate tracking AGN activity does not imply 
that these processes occur in tandem in all individual 
galaxies \citep{treister2013}.
 
While these relations imply that the evolution of a galaxy
and its central black hole are intertwined, their exact origin remains uncertain. Several
mechanisms have been proposed. One possibility is that the evolution of a galaxy regulates
the growth of its central black hole by determining the amount of gas that finally reaches
the black hole \citep{booth2010, booth2011}. However, a SMBH can also regulate the evolution of its
host galaxy through energy input via AGN feedback, which may be able to suppress star formation
\citep{maiolino2012}. \citet{dubois2012} have run recent cosmological simulations including
SMBH growth via accretion and mergers, as well as outflows and heating from AGN feedback,
and conclude that AGN feedback is necessary to expel gas and suppress star formation.
They also argue that AGN feedback is able to transform late-type galaxies
into early-types and explain galaxy scaling relations such as the fundamental
plane \citep{dubois2013}. Similarly, \citet{choi2013} simulate mergers of galaxies
including radiative and mechanical AGN feedback and are able to recover
the $M_{\bullet} - \sigma$ relation as well as reproduce the X-ray luminosities
of galactic halo gas seen in observations, although they find that AGN feedback is insufficient
to explain galaxy size evolution.
 
It is not known which of the former two processes is dominant. A third explanation of the scaling relations
is that some other factor --- for instance, the total gas reservoir --- regulates both the growth of the SMBH
and the evolution of the galaxy. \citet{bournaud2012} observed a correlation between
giant clumps of gas and stars, which are indicative of violent disk instabilities, and AGN activity at $z \sim 0.7$. 
They propose that the evolution of disk instabilities, which cause gas inflow to the central bulge and SMBH,
could produce the observed correlation between star formation and AGN activity \citep{bournaud2011}. 

Finally, a purely statistical explanation has been proposed for the existence of these observed correlations
based on the idea that repeated mergers of galaxies, which also lead to the eventual mergers
of their central black holes, can cause the SMBHs and their hosts to have correlated masses as
a consequence of the central limit theorem \citep{h2010}. One or a combination of the above
mechanisms could be responsible for the observed scaling relations between SMBHs and their
hosts. The argument based on the central limit theorem has recently been gaining strength as
the evidence accumulates from both observations \citep{dokkum2008} and theory \citep{oser2010}
that the most massive galaxies grow their stellar components substantially via minor mergers
during recent cosmic epochs. This is especially important for massive
BCGs (brightest cluster galaxies) in rich clusters \citep{hausman1978, lin2013}. Evidence that BCGs have 
grown via mergers has been reinforced by a new statistical test \citep{lin2010}. 
Recent observations of early-type galaxies showing that the color-magnitude and color-size relations change
slope for  $M_{*} > 2\times10^{11}$, but the color-$\sigma$ relation
stays constant, suggest that major dry mergers are significant in the evolution of such massive
galaxies \citep{bernardi2011a, bernardi2011b}. The impact of the increased
 frequency of mergers for the assembly of the stellar component suggests that they might play an important 
role in black hole growth as well at late times.

It is widely believed that black hole growth occurs primarily via accretion or merger triggered accretion episodes
over cosmic time, although secular evolution driven by stellar evolutionary processes
also appears to be important \citep{ciotti1997, ostriker2010}, especially for $z < 2$. In massive galaxies, 
which have undergone multiple mergers, it is believed that most of the SMBH growth occurs in short accretion episodes
fueled by gas flowing into the central region due to the merger or galactic cooling flows. SMBHs hosted in small galaxies that 
have not undergone any major mergers must, however, be supplied with gas through some 
different mechanism, such as dynamical relaxation and perhaps secular evolution processes \citep{sesana2011}.
  While these may indeed be the primary channel of growth during early times, it is
becoming evident that the actual merger of black holes might contribute appreciably to the
final mass inventory of the most massive black holes at low redshifts. This is 
suggested by the current evidence that minor mergers are a significant component of the 
late-time stellar mass growth of the most massive galaxies.

In this paper, we examine how gas inflows, mergers, and the large-scale environment determine black hole
growth. To this end we track the growth histories of black holes in a typical over-dense
cluster environment and an under-dense void environment. In particular, we know that the merger
history in these environments at late times is divergent. Therefore, in this work we focus
on quantifying the role that mergers play in the mass assembly history at late times. This
has important consequences for the expected gravitational radiation from such events at low
redshifts as well the observational implications for the number of wandering black holes.
Since BH-BH mergers will be accompanied by the gravitational slingshot effect \citep{peres1962, bekenstein1973,
fitchett1984}, late time mergers may eject a significant amount of mass in BHs from merged galaxies.
An inventory of BHs at the present epoch will, of course, not count this mass and 
thus will underestimate the $z=0$ cosmic mass density locked up in SMBHs. Furthermore, 
we expect that some SMBHs at any given time will be orbiting in the outskirts of the
galactic potential --- as the result of a lower velocity gravitational wave recoil or 
a recent merger of its host with a more massive galaxy. These orbiting BHs will also
 not be found in observations of galactic centers. The correction to the SMBH mass density from
these two populations, which we estimate here, implies that the Soltan argument \citep{s1982} will 
necessarily overestimate the efficiency of energy output from accreting BHs. This is because the energy output
from accretion that is directly observed is incommensurate with the SMBH cosmic mass density estimated only
from the census of BHs in galactic centers. Neglecting the ``unmerged'' SMBH population 
(i.e., ejected and orbiting BHs) introduces errors
in estimates of the inferred accretion efficiency for the population of BHs.

The outline of the paper is as follows: we first describe the key aspects of the problem
that are tackled here --- the role of late time mergers --- in \S\ref{s2}. In \S\ref{s3}, the methodology
and cosmological simulations used are detailed, followed by a discussion of previous
work on this topic and our results in \S\ref{s41}.
We conclude with a discussion of the implications and observational consequences
of accounting for the role of mergers at late times on our current understanding of 
black hole growth.

\section{Exploring the role of low-redshift mergers}
\label{s2}

In this paper, we examine the consequences of mergers for the mass assembly history of SMBHs
at late times and the observational consequences thereof. The ``Soltan argument''
\citep{s1982} relates the observed luminosity density of AGN over time to the local
mass density in SMBHs. The energy radiated by AGN is proportional to the rate
at which they gain mass via accretion --- $L = \epsilon \dot{M}_{acc} c^{2}$, where $\epsilon$
is the radiative efficiency factor. Since the total mass in SMBHs at $z = 0$ is the
integral of the mass they have accreted over all time, one can relate the observed luminosity density
of AGN to the total mass density of SMBHs at $z = 0$ \citep{shen2009}:
\begin{equation}
\label{soltaneq}
\rho_{\bullet} \approx \int_{0}^{\infty} \frac{\der t}{\der z} \der z \int_{0}^{\infty} \frac{(1 - \epsilon)L}{\epsilon c^{2}}
\Phi(L, z)\der L
\end{equation}
where $\Phi(L, z)$ is the AGN bolometric luminosity function per $L$ interval.
The local SMBH mass density $\rho_{\bullet}$ has been found to be consistent
with the observed luminosity density of AGN if these AGN
have a mean mass-to-energy conversion efficiency of $\epsilon \simeq 0.1$ \citep{yutrem2002, marconi2004}.

While SMBHs can grow through BH-BH mergers as well as accretion, such mergers 
can only alter the shape of the SMBH mass function, having no significant effect on the total mass density
of SMBHs \citep{menou2004,shen2009, shankar2009}. There is a small mass loss during mergers 
due to gravitational wave radiation, which is a negligible correction to the Soltan argument, but is important
to know accurately for gravitational wave detection experiments \citep{rajagopal1995, jaffe2003}.

Many of the existing calculations of the energy emitted in gravitational 
waves during BH-BH mergers use Monte-Carlo dark matter merger trees (e.g.,
\citealt{wyithe2004, sesana2004, enoki2004}); however, these approaches
do not take into careful account the enhanced merger rates in cluster environments where the
most massive galaxies live \citep{volonteri2013}. Some more recent works 
\citep{sesana2009, kocsis2011, ravi2012} have 
used dark matter merger trees based on the \textit{Millennium} Simulation \citep{springel2005}.
The dynamical influence of the galaxy
stellar mass is not taken into account in dark matter-based merger trees, but it
can greatly shorten the time for two galaxies to merge. Assuming an NFW density profile
\citep{nfw} for the dark matter halo, the dynamical friction time 
for a galaxy in a subhalo to merge with the central galaxy has the form
 $t_{\mathrm{DF}} \propto (1+M_{*}/M_{\mathrm{DM}})^{-9}$ \citep{mcwilliams2012b}, where
$M_{*}$ is the stellar mass and $M_{\mathrm{DM}}$ is the halo mass of the tidally limited subhalo.
Hydrodynamical cosmological simulations can 
account for the dynamics of both the dark and baryonic matter.

Other recent works have estimated the gravitational wave emission
based on observational constraints. \citet{mcwilliams2012} calculated the 
gravitational wave strain for $z \le 1$ using observed galaxy mass functions
combined with observed BH-galaxy scaling relations, assuming that SMBH
growth at these times occurs purely via BH-BH mergers.
\citet{sesana2013, sesana2012} used observed
galaxy stellar mass functions and pair fractions, combined with galaxy merger
timescales from the \textit{Millennium} Simulation and hydrodynamical simulations
of galaxy mergers, to calculate galaxy merger rates. These rates were then 
combined with observed BH-galaxy scaling relations to obtain the expected strain.

Not all galactic mergers will result in black hole mergers, due to a finite dynamical
friction time for the black hole to sink to the center of the galactic potential.  There also
exists the ``final parsec problem'' (see e.g., \citealt{mm2003}), which is shorthand for the
physical difficulties in bridging the gap between the binary separation reached by dynamical
friction between the SMBH and the ambient stellar population and the much smaller separation
at which gravitational radiation takes over as the primary angular momentum loss mechanism for the 
BH binary. We will not address this problem in this paper, but will assume that black holes effectively 
merge at galactic centers on the dynamical friction timescale. 

The merger of two SMBHs can also impart a linear momentum to the resulting
post-merger SMBH due to gravitational wave radiation \citep{peres1962, bekenstein1973,
fitchett1984}. This can cause the SMBH to be displaced from the center 
of the galaxy or ejected entirely; we discuss this in more detail in \S \ref{sec:recoils}.

Calculation of the mean mass-to-energy conversion efficiency of SMBHs using the Soltan
argument requires an accurate census of the local mass density of SMBHs $\rho_{\bullet}$.
Because we observe SMBHs only at the centers of galaxies,
the existence of SMBHs outside of galactic centers due to the mechanisms described above
implies that some of the local SMBH mass is unaccounted for by observations,
and thus that the radiative efficiency $\epsilon$ derived from the Soltan argument
is an overestimate. In this paper, we estimate the fraction of total SMBH mass
that was accreted throughout cosmic time but is not currently observable
in galactic centers, which must be added to the left side of Equation \ref{soltaneq}
in order to obtain the correct mean $\epsilon$. Unlike some
semi-analytic works \citep{shankar2009, shen2009, shankar2013} that
use observed AGN luminosity functions to constrain SMBH evolution, the merger trees we
use in this work are entirely derived from hydrodynamical cosmological simulations and
are thus independent of the assumed radiative efficiency.

Taking galaxy and black hole mergers as well as gravitational wave recoils 
explicitly into account will alter the observational consequences at low redshifts and
that is precisely what we explore in detail in this paper.
We use recent cosmological simulations \citep{c2011, cen2011, cen2012b, cen2012a, cen2013}
 to study the relevance of different processes on the growth of SMBHs and their implications for the setting
up of the various observed scaling relations in two significantly different environments. In our book-keeping
we keep track of the total mass in black holes in the following four categories at redshifts between $z = 4$ and 0:

\begin{itemize} 
\item{The mass in central black holes acquired from accretion of gaseous matter.} \\
\item{The mass in central black holes acquired from mergers with smaller black holes.} \\
\item{The mass in orbiting black holes that have not yet fallen to the center of the galactic potential.}\\
 \item{The mass in ejected black holes that have been kicked out due to gravitational radiation recoil.} 
\end{itemize}

The final two categories will be assessed as corrections to the normal Soltan-type arguments
for estimating the revised efficiency of BH-associated energy generation.

\section{Methodology and description of cosmological simulations}
\label{s3}

As the basis for our analysis of SMBH evolution, we use the large-scale
hydrodynamical galaxy simulations of \citet{c2011, cen2011, cen2012b, cen2012a, cen2013}. Detailed
descriptions of the simulations can be found in the papers referenced above;
we provide a brief overview in this section.  We use galactic (not dark matter based) merger trees
derived from these simulations as described in \citet{lack2012}. The
simulations are performed with the AMR (Adaptive Mesh Refinement)
Eulerian hydrodynamics code Enzo \citep{bry1999, shea2004, joung2009}.
They consist of a low-resolution box of 120 $h^{-1}$ Mpc on a side,
and two high-resolution regions within this box, one containing a
cluster with mass $\sim3\times10^{14}$ $M_{\odot}$, and the other a
void. These represent $+1.8\sigma$ and $-1.0\sigma$ fluctuations in the
cosmic density field, respectively. These two extreme regions bracket the cosmic average 
environment. The cluster region box has dimensions
$21\times24\times20$ $h^{-3}$ Mpc$^{3}$, and the void box has dimensions
$31\times31\times35$ $h^{-3}$ Mpc$^{3}$. The dark matter particle mass
is $1.07\times10^{8}$ $h^{-1}M_{\odot}$, while the stellar particle mass
is generally around $10^{6} M_{\odot}$. The resolution in the cluster
and void regions is always better than 460 $h^{-1}$ pc physical.
Galaxies in the simulation box are identified by using the HOP algorithm on stellar particles
\citep{eisenstein1999}.

The simulation includes prescriptions for UV background \citep{haardt1996}, shielding from UV
radiation by neutral hydrogen \citep{cen2005}, metallicity-dependent radiative cooling \citep{cen1995},
formation of star particles from gas \citep{cenostriker1992}, and supernovae feedback \citep{cen2005}. It
does not include feedback from AGN, which may be partly the reason that the
largest galaxies in the simulation box have too many stars in comparison with
observed relations. This is a well known problem in hydrodynamical simulations \citep{oser2010, guo2010} 
and one that does exist in our version. Broad agreement is otherwise found between the simulation
results and observations \citep{c2011}.

The simulations use the following cosmological parameters, consistent
with \citet{komatsu2010}: $\Omega_{M} = 0.28, \Omega_{b} = 0.046, \Omega_{\Lambda} = 0.72,
\sigma_{8} = 0.82, H_{0} = 100 h^{-1}$ Mpc$^{-1} = 70$ km s$^{-1}$ Mpc$^{-1}$,
and $n = 0.96$. These are also the values we adopt throughout this paper in our calculations.

A merger tree is created from this simulation. In the cluster box, there
are 38 redshift slices from $z = 4$ to $z = 0$ with $\Delta z = 0.05$
for $z < 1.35$, and slices at $z =$ 1.5, 1.6, 1.75, 1.9, 2.0, 2.2, 2.5,
2.8, 3.1, and 4. In the void box, there are 14 redshift slices between
$z = 0$ and $z=4$, at $z =$ 0, 0.05, 0.15, 0.2, 0.4, 0.5, 0.6, 0.8, 1.0,
1.6, 1.9, 2.5, 3.1 and 4 \citep{lack2012}. 

Since our focus is on massive galaxies, and due to the resolution limit of the simulation, 
our merger tree contains only galaxies with
stellar masses $M_{*} > 10^{9} M_{\odot}$. In the void box, 
the grouping algorithm identifies some groupings of particles as galaxies 
that do not exist in later redshift slices. For our purposes, we ignore any groupings of particles 
that do not have a descendant at $z=0$.

We first scale down the stellar masses of the galaxies in this merger tree 
to account for the overproduction of stellar mass in our simulation.
We then seed these galaxies with central black holes whose evolution we trace
based on the evolution of their host galaxies. Further details are provided below.

\subsection{Scaling of Cluster and Void Boxes}
\label{scaling}

Our simulation considerably overproduces stellar mass compared to the amount of dark matter
present. This is a common problem in cosmological simulations \citep{oser2010, guo2010}.
Because observations find that the masses of central black holes scale with 
the stellar content of their host galaxies, using the galaxy masses as-is would lead us to greatly overestimate
the amount of mass contained in black holes. We attempt to allow for this
by scaling down the stellar masses of our galaxies by a constant factor
in proportion to the excess of star formation efficiency as normalized by current observations.

In the cluster, the stellar mass is $3\times 10^{13} M_{\odot}$ within $r_{200}$, the radius
within which the mean density is equal to 200 times the critical density. The dark matter
mass within this same radius is $3\times10^{14} M_{\odot}$ \citep{lack2012}. 
This implies a stellar to dark matter ratio of 0.1 within the virial radius.
Comparing this to the stellar-halo mass relation found by 
\citet{leauthaud2012} using weak lensing and halo occupation
distribution methods, combined with the fraction of halo mass that is in gas given by
\citet{pratt2009}, one finds that our simulation overproduces stars
for a cluster of this mass by a factor of roughly 4 to 6.
 Determinations of the stellar-halo mass relation found 
by matching simulated dark matter halos to observed galaxy mass functions
 \citep{guo2010, behroozi2010, moster2013}
find somewhat lower values than \citet{leauthaud2012}, implying 
an even larger excess for our simulation. Recent observational
determinations of low redshift cluster stellar masses using WISE and 2MASS 
 and halo masses using \textit{Chandra} find values of the stellar to dark matter ratio in the range
$\approx 0.01 - 0.03$ \citep{lin2012} within $r_{500}$. This corresponds to an overproduction
of stellar mass by a factor of approximately 3 to 10. Given these results, we 
assume the stellar to dark matter mass ratio within the virial radius of the cluster
to be 0.02, implying a scaling factor for galaxies in the cluster of 0.2.
We apply this scaling factor to all the galaxies in the cluster box (not only those
within the virial radius), within which the initial stellar to dark
matter mass ratio is 0.06. Thus, after scaling, this ratio is
0.012 for the entire cluster box. However, because overcooling is 
less of a problem in the outskirts of the cluster region than the center,
this stellar to dark matter mass ratio may be somewhat too low.

Due to the lack of observational data for the star formation efficiency
specifically in the void, we assume arbitrarily that the stellar to
dark matter mass ratio in the void is half of the mean ratio of
the universe. We calculate the mean universal value using the 
$z = 0$ stellar mass density obtained by
\citet{bernardi2013} for a Sersic + Exponential
fit to galaxy luminosity profiles: $\rho_{*} = 3.30 \times 10^{8} M_{\odot}$ Mpc$^{-3}$.
Although this value is for galaxies with $M_{*} > 10^{9} M_{\odot}$, galaxies
with lower stellar masses are expected to contribute a negligible
amount of the total stellar mass in the universe \citep{brinchmann2004}.
This results in a mean stellar to dark matter mass ratio of 0.0104, and
thus a ratio of 0.0052 for the void. The stellar to dark matter ratio
for the entire void box in our simulation is 0.012 \citep{lack2012},
so we scale the stellar masses of the galaxies in the void by a factor of 0.43 
in the same manner that we do for the cluster.

It should be noted that these scalings for the void and cluster box
are simple and do not reflect the more complex trends in stellar mass overproduction
in our simulation. A discussion of the effect of our choices for the scaling 
can be found in \S \ref{s4}.

When evolving our SMBHs, we grow them proportionally to the
bulge mass in their host galaxy (see \S\ref{sec:evolution} below for more detail),
in agreement with observations. We set the proportionality constant so that
the resulting SMBH mass per unit dark matter matches that observed in cluster
and void regions in the real universe. To accomplish this, we 
first estimate the fraction of stellar mass in bulges in cluster and void regions.
We use a mean bulge-to-total mass ratio as function of stellar mass of 
\begin{equation}
\label{bulge}
\frac{M_{\mathrm{*,bulge}}}{M_{*}} = \left[1 + \left(\frac{M_{*}}{2\times10^{10} M_{\odot}}\right)^{-0.58}\right]^{-1}.
\end{equation}
This is consistent with the bulge mass fractions obtained from a sample of
$\sim 660,000$ SDSS galaxies in \citet{mendel2013} (aside from a drop
in bulge fraction at $M_{*} \gtrsim 10^{11} M_{\odot}$ 
not seen in previous studies, e.g. \citealt{gadotti2009, cibinel2013}).
We obtain galaxy stellar mass functions for
high and low density regions of the universe from \citet{bolzonella2010}.
Combining these with the bulge-to-total mass ratio as a function of stellar mass 
results in a bulge stellar mass fraction 
of $71\%$ in the cluster and $51\%$ in the void. We thus find a bulge stellar mass to
dark matter mass ratio of $8.52\times10^{-3}$ in the cluster and 
$2.65\times10^{-3}$ in the void. We note that we normalize the entire
cluster box as though it had the stellar mass function of a high density region,
which may not be the case for the outskirts of the box. Thus, while the stellar to 
dark matter ratio in the cluster box is likely somewhat underestimated, the fraction
of stellar mass in bulges is likely somewhat overestimated.

Using these values for the bulge mass per unit dark matter, we derive the 
SMBH mass per unit dark matter. We use the observed $M_{\bullet}-M_{\mathrm{bulge}}$
relation from \citet{mcc2012}, which implies a mean black hole to
bulge mass ratio of $\approx 4\times10^{-3}$ when taking into account
the scatter in the relation. We assume this value for both the cluster and void regions
and obtain a SMBH mass per unit dark matter mass of $3.41\times10^{-5}$ in the
cluster and $1.06\times10^{-5}$ in the void. We grow the SMBHs in proportion
to their host bulges with a distribution such that we obtain these values 
at $z=0$ in the cluster and void boxes; a full description of
the SMBH growth prescriptions is given in \S\ref{sec:evolution}.

The cluster and void boxes in the cosmological simulation we use 
are +1.8$\sigma$ and -1$\sigma$ fluctuations in the
cosmic density field, respectively, and were chosen so as to bracket the
``global average'' of various physical quantities \citep{c2011}. To approximate
this global average, we combine the rescaled quantities from the void and cluster boxes
in a weighted average. We choose weights such that the fraction
of stellar mass per unit dark matter mass matches $\rho_{*}/\rho_{\mathrm{DM}}$ in 
the local universe at $z = 0$. We use the local stellar mass density
of $3.3\times10^{8} M_{\odot}$ Mpc$^{-3}$ for galaxies with $M_{*} > 10^{9} M_{\odot}$
found in \citet{bernardi2013}, resulting 
in a stellar to dark matter mass ratio of 0.0104. We note that the value of the local stellar mass
density used also carries significant uncertainty, and depends on both
the assumed initial mass function and stellar mass-to-light
ratio of galaxies, as well as the photometry used to 
obtain the galaxy luminosity functions from which the stellar mass
functions are derived \citep{bernardi2013}. To match
this stellar to dark matter mass ratio at $z=0$ requires that we 
approximate the universe as $77\%$ cluster
per unit dark matter mass and $23\%$ void. We use these same weights at 
all redshifts.

We multiply by the dark matter density of the universe 
where necessary to obtain various number and mass densities cited throughout our
paper. The black hole to dark matter mass ratios derived above
for the universe correspond to a universal SMBH mass density
of $9.1\times10^{5} M_{\odot}/$ Mpc$^{-3}$. This value is higher than
previous estimates, which are in the range $3-5.5\times10^{5} M_{\odot}/$ Mpc$^{-3}$ 
\citep{shankar2009}. The main source of the difference is
the BH to bulge mass ratio implied by the observations of \cite{mcc2012},
which is higher than that used to obtain previous
 estimates of the SMBH mass density (see also \citealt{kormendyreview}).

We run multiple realizations of our randomized model and use them to
compute one-sigma errors on the results we obtain. However, it should be noted 
that although our black hole growth is modeled with random scatter, we
are always using the same galaxy merger trees obtained from the simulations of
\citet{c2011}.

\subsection{Black Hole Evolution Prescriptions}
\label{sec:evolution}

The masses of central supermassive black holes are known to correlate well with the masses
of host elliptical galaxies and the bulge component of host spirals. Correlation with the disk
of spiral galaxies seems to be weaker or nonexistent 
(e.g., \citealt{kormendyreview} and references therein; see however \citealt{lasker2014}).
Thus we take the growth of our SMBHs to be proportional to that
of the bulge stellar mass in our galaxies.
However, our simulation does not have the resolution to distinguish bulges from disks, so we assume that 
the bulge mass of galaxies is a function of their stellar mass following Equation \ref{bulge}.

For simplicity we assume the same relation between bulge and total stellar mass at all redshifts,
although it is expected from models that the bulge mass fraction at fixed
stellar mass should increase with decreasing redshift \citep{somerville2012, guo2013, avila2013}.
Recent observations also show an increase in bulge stellar mass fraction between $z = 2.5$
and $z = 0.5$, although the increase is slight \citep{lang2014}. Since
we do not take such evolution into account, it is likely that a larger
fraction of SMBH growth takes place at early times in our model than in
the real universe. We grow our SMBHs such that the black hole
mass to dark matter mass ratio at $z = 0$ is correct by design (\S \ref{scaling}),
meaning that the SMBHs we obtain at early times may be overly massive.

We place a seed black hole in any galaxy that reaches our lower mass limit
at any redshift; this is $M_{*} = 10^{9} M_{\odot}$ for the original simulation masses,
and thus $2\times10^{8} M_{\odot}$ for the cluster and $4.3\times10^{8} M_{\odot}$ for the 
void in terms of rescaled masses. The seed black hole is taken
to have mass $nM_{*, \mathrm{bulge}}$, where $n$ is selected from a log-normal distribution with
median value such that the black hole to dark matter mass ratio at $z=0$ matches observations (\S\ref{scaling}),
and with scatter as described further below.
The median proportionality factor is $1.50 \times 10^{-3}$ for the cluster and $1.25 \times 10^{-3}$ for the void.
This is consistent with the common assumption that of the gas added to (and retained
by) galaxies, approximately one part in a thousand is accreted onto the central black hole,
with most of the remainder transformed into stars \citep{lihaiman2007}. We note
 that this assignment of BH seed masses as early as $z = 4$ pre-supposes the existence of a scaling relation 
at this epoch akin to what is empirically measured at $z = 0$.

SMBHs are then allowed to grow through BH-BH mergers and accretion of gaseous material from the galaxy.
We calculate the accreted and merged mass by considering all parent black holes of a single black hole to 
be ``merged mass'' except for the most massive one. We adopt a simple prescription for accretion in which
the mass accreted by the central black hole in each redshift slice is proportional to 
the stellar mass formed in its host galaxy in that redshift slice times
the bulge fraction of the host galaxy, with some scatter in the assumed proportionality factor. 
A physical mechanism proposed for such a proportionality is that gas infall onto the SMBH is 
caused by the gas drag due to stellar radiation, which in turn is roughly proportional
to the star formation rate \citep{umemura2001,kawakatu2002, kawakatu2003, granato2004}.
Because stars are sometimes ejected from galaxies in the simulation, it is possible for the total stellar
mass of a galaxy to occasionally decrease. If this is the case, we assume zero accretion onto the
black hole in that redshift slice. Otherwise, the mass accreted by the black hole is taken to
be $n^{\prime} \Delta M_{*, \mathrm{bulge}}$, where $\Delta M_{*, \mathrm{bulge}}$ is the change in bulge stellar
mass due to star formation in that redshift slice, and $n^{\prime}$ is a proportionality factor
chosen from the same log-normal distribution as the proportionality factor for the seeding prescription.
We choose such a prescription for accretion due to the 
observed correlation between AGN activity and galactic star formation rate
\citep{heckman2004, merloni2004, brusa2009, shankar2009}.

The proportionality factors for the seed masses and BH accretion rates
are chosen from the same log-normal distribution, with median values
$1.50 \times 10^{-3}$ for the cluster and $1.25 \times 10^{-3}$ for the void as described
above. We adjust the scatter of the distribution so as to produce an
intrinsic scatter of $\sim 0.35$ in the cluster 
$M_{\bullet}-M_{*, \mathrm{bulge}}$ relation at $z=0$, similar to observed
values reported in the literature \citep{sani2011, mcc2012}.

When a smaller galaxy merges into a larger one, we calculate the dynamical time for the black hole
from the smaller galaxy to move to the center of the newly formed merged galaxy. We 
calculate the dynamical friction time as: 
\begin{equation}
\label{bt}
\!t_{\mathrm{DF}}{=}\frac{19\,\textrm{Gyr}}
{\ln(1 {+} M_{*}/M_{\bullet})}\left(\frac{R_{e}}{5\, \textrm{kpc}}\right)^{2} \frac{\sigma}{200 \,
\textrm{km/s}}\frac{10^{8}\, M_{\odot}}{M_{\bullet}} f_{e}.
\end{equation}
The equation is that for a circular orbit from
 \citet{b1987} and is corrected by a factor $f_{e}$ for the ellipticity
of the orbit. Following the analysis of \citet{gnedin2013}, who
find that $f_{e} \simeq 0.39-0.76$ for satellite halos
merging with a central galaxy, we adopt $f_{e} = 0.5$. 
Although choosing a different value of $f_{e}$ will obviously
alter the number of orbiting BHs and the number of BH-BH mergers
by increasing or decreasing the dynamical friction time, 
the difference is not large enough to significantly affect our general
results. For example, adopting even the extreme case of
purely circular orbits ($f_{e} = 1$)
increases the fraction of mass in orbiting BHs at $z=0$ from $7\%$ to $10\%$,
while reducing the mass contribution from BH-BH mergers in massive
BHs from $\approx24\%$ to $\approx20\%$.

We obtain values for the effective radius $R_{e}$ and velocity dispersion $\sigma$ of our simulated
galaxies by using observed fits to the stellar mass from SDSS data at $z=0$ \citep{n2009}. Since
the relations between galactic mass, radius, and velocity dispersion are known to
evolve with redshift, we take $R_{e} \propto (1+z)^{-0.98}$ at fixed mass based on the
observations compiled in \citet{vanderwel2008} and \citet{mclure2012}. Similarly, we
take $\sigma \propto (1+z)^{0.47}$ based on observations of the mass
fundamental plane out to $z \sim 2$ \citep{bezanson2013}. 
Also, both the size and velocity dispersion evolution are consistent with the 
recent observations of \citet{belli2013} for $0.9 < z < 1.6$. Thus our final scaling relations are:
\begin{equation}
\label{eqn2}
R_{e} = 2.5 \, \textrm{kpc}
\left(\frac{M_{*}}{10^{11}\,M_{\odot}}\right)^{0.73}(1+z)^{-0.98},
\end{equation}
\begin{equation}\label{eqn1}
\sigma(R_{e}) = 190 \, \textrm{km/s} \
\left(\frac{M_{*}}{10^{11}\,M_{\odot}}\right)^{0.2}(1+z)^{0.47}.
\end{equation}

If a galaxy is involved in another merger before a satellite black hole has merged with the
central one, we recalculate the dynamical friction time. If the satellite black hole is in the
more massive galaxy, we take the new dynamical friction time to be the smaller of the remaining
time to black hole merger and the dynamical friction time that would be calculated for the
post-merger galaxy. For a less massive galaxy merging into a more massive one, its satellite
and central black holes are taken to have a dynamical friction time calculated for the new
post-merger galaxy. 

If a black hole is ejected or displaced from the center of a galaxy due to a 
gravitational wave recoil (see \S\ref{sec:recoils} below), the galaxy may have no
central SMBH. If a galaxy lacking a central BH produces an additional 
$10^{9} M_{\odot}$ in unscaled galactic star formation (i.e., $2\times10^{8} M_{\odot}$ in
the cluster and $4.3\times 10^{8} M_{\odot}$ in the void after scaling), 
we seed it again with a black hole with mass proportional to the (scaled) mass
in stars formed times the bulge mass fraction, where
the proportionality constant is chosen from the same log-normal distribution as for the
seeds and accreted mass described earlier.

\subsection{Low redshift merging and gravitational radiation}
\label{sec:merging}

We calculate the gravitational radiation luminosity expected based on the BH-BH
mergers that occur in our simulation. This energy is equivalent to the mass
lost from the black hole pair as they are merging. For each BH-BH merger, we apply an approximation
for the energy emitted in gravitational waves from \citet{barausse2012b}: 
\begin{flalign}
\label{gwfirst}
\frac{E_{rad}}{M} = &[1-\tilde{E}_{\textrm{ISCO}}(\tilde{a})]\nu + \notag \\ & 4\nu^{2}[4p_{0}
+ 16p_{1}\tilde{a}(\tilde{a}+1) + \tilde{E}_{\textrm{ISCO}}(\tilde{a} - 1)].
\end{flalign}
Here $M \equiv m_{1} + m_{2}$ is the total mass of the two black holes and
 $\nu \equiv m_{1}m_{2}/M^{2} $ is the symmetric mass ratio. The constants
$p_{0}$ and $p_{1}$ come from a polynomial fit to the emitted energy, and have the values: 
\begin{equation} 
p_{0} = 0.04827 \pm 0.00039, \quad p_{1} = 0.01707 \pm 0.00032.  
\end{equation} 
Here $\tilde{E}_{\textrm{ISCO}}$ is the energy per unit
mass at the innermost stable circular orbit and is given by:
\begin{equation} 
\tilde{E}_{\textrm{ISCO}}(\tilde{a})
= \sqrt{1-\frac{2}{3\tilde{r}^{eq}_{\textrm{ISCO}}}(\tilde{a})}, 
\end{equation}
where
\begin{flalign}
 \tilde{r}^{eq}_{\textrm{ISCO}}(\tilde{a}) = & 3 + Z_{2} \notag
\\ & - \mathrm{sign}(\tilde{a}) \sqrt{(3-Z_{1})(3+Z_{1}+2Z_{2})}, \notag
\end{flalign}
\begin{flalign}
 Z_{1} = & 1 + (1-\tilde{a}^{2})^{1/3}\left[(1+\tilde{a})^{1/3} +
(1-\tilde{a})^{1/3}\right], \notag
\\ Z_{2} = & \sqrt{3\tilde{a}^{2}+Z_{1}^{2}}. 
 \end{flalign}
In all the above equations, $\tilde{a}$ is defined as 
\begin{equation}
\label{gwlast} 
\tilde{a} \equiv \frac{\boldsymbol{\hat{L}}\cdot(\boldsymbol{S}_{1} + \boldsymbol{S}_{2})}{M^{2}} =
\frac{|\boldsymbol{a}_{1}|\cos{\beta}+q^{2}|\boldsymbol{a}_{2}|\cos{\gamma}}{(1+q)^{2}},
\end{equation} 
where $q \equiv m_{1}/m_{2} < 1$ is the mass ratio of the two black holes,
$|\boldsymbol{a}_{1}|$ and $|\boldsymbol{a}_{2}|$ are the spin magnitudes, and $\beta$ and
$\gamma$ are the angles between the orbital angular momentum unit vector $\boldsymbol{\hat{L}}$
and the spins of the first and second black hole, respectively. By construction, this formula 
is accurate in both the test-particle limit and for equal-mass
binaries. 

A number of works have investigated the expected distributions of the
magnitude and orientation of SMBH spin vectors, using both semi-analytic models 
(e.g., \citealt{volonteri2005, barausse2012a,
volonteri2012}) and hydrodynamical simulations (e.g., \citealt{dotti2012, dubois2013b}).
 SMBHs are spun up by coherent
gas accretion and spun down by chaotic accretion \citep{volonteri2012, barausse2012a}. 
In a gas-poor merger, the spin directions of the merging BHs are
expected to be randomly distributed \citep{bogdanovic2007, barausse2012a}, and
the merger remnant tends to be spun down compared to the initial central BH \citep{volonteri2012}.
Alternatively, in a gas-rich merger, gas accretion can exert torques that align the spins
of both BHs with the angular momentum of the gas accretion disk to within $10^{\circ} - 30^{\circ}$
\citep{bogdanovic2007, dotti2009}, which creates a spun-up merger remnant with the
same spin direction as the two inspiraling BHs \citep{dotti2009}.
However, quantitative predictions of trends in black hole spin with SMBH mass and redshift
tend to vary between models and depend on the assumptions made.
For the main results presented in this paper,
we make the most ``simple'' possible assumption by selecting spin parameter
values from a random uniform distribution between 0 and 1, 
and taking the direction of the spins to be uniformly randomly
oriented over a sphere. This is similar to the assumptions made by
\citet{schnittman2007} when calculating the expected recoil velocity
distribution of SMBHs. If we assume the opposite extreme ---
that all merging BHs are aligned with the orbital angular momentum to within $10^{\circ}$
--- the only result that is significantly affected is the number of 
 ejected BHs. We describe the effect of assuming 
partially aligned BH spins in \S\ref{s4}.

\subsubsection{Gravitational Wave Recoils}
\label{sec:recoils}

When two orbiting supermassive black holes merge, the 
gravitational radiation produced can impart a linear (``slingshot'') momentum to the
SMBH resulting from the merger \citep{peres1962, bekenstein1973,
fitchett1984}. Recent numerical simulations of general relativity
 show that such ``kicks'' can in some cases be large
enough to exceed the escape velocity of the host galaxy and
eject the resulting SMBH \citep{herrmann2007, koppitz2007,
campanelli2007a, campanelli2007b, lousto2011}. Gravitational wave recoils 
with velocities insufficient to eject the SMBH can still displace it from
the center of the galaxy, to which it may return via dynamical friction.

We account for the effects of gravitational wave ``kicks'' on our 
population of SMBHs, including both ejections and displacements
from the center of the galaxy. We use the 
fitting formula from \citet{lousto2012}
for the velocity imparted to an SMBH resulting from a merger:
\begin{flalign}
& \boldsymbol{V}_{\textrm{recoil}}(q, \boldsymbol{\alpha})
 = v_{m}\boldsymbol{\hat{e}}_{1} + v_{\perp}(\cos\xi\boldsymbol{\hat{e}}_{1}+\sin\xi\boldsymbol{\hat{e}}_{2})
+v_{\parallel}\boldsymbol{\hat{n}}_{\parallel}, \notag \\
& v_{m} = A_{m}\frac{\eta^{2}(1-q)}{(1+q)}[1+B_{m}\eta], \notag \\
& v_{\perp} = H\frac{\eta^{2}}{(1+q)}\left[(\alpha_{2}^{\parallel}-q\alpha_{1}^{\parallel})\right], \notag \\
& v_{\parallel} = 16\eta^{2}/(1+q)\left[V_{1,1}+V_{A}\tilde{S}_{\parallel}+V_{B}\tilde{S}_{\parallel}^{2}+V_{C}\tilde{S}_{\parallel}^{3}\right] \notag \\
& \times |\boldsymbol{\alpha}_{2}^{\perp}-q\boldsymbol{\alpha}_{1}^{\perp}|\cos(\phi_{\Delta}-\phi_{1}).
\end{flalign}
Here $\eta = q/(1+q)^{2}$, where $q = m_{1}/m_{2}$ is the mass ratio of the smaller
to larger black hole, $\boldsymbol{\alpha}_{i} = \boldsymbol{S}_{i}/m_{i}^{2}$ is the dimensionless
spin of black hole $i$, $\parallel$ and $\perp$ refer to components
parallel and perpendicular to the orbital angular momentum, respectively,
$\boldsymbol{\hat{e}}_{1}$ and $\boldsymbol{\hat{e}}_{2}$ are orthogonal unit vectors in the orbital plane,
$\xi$ is the angle between the unequal mass and spin contributions to the
recoil velocity in the orbital plane, and $\boldsymbol{\tilde{S}} = 2(\boldsymbol{\alpha}_{2}+q^{2}\boldsymbol{\alpha}_{1})/(1+q)^{2}$.
$\phi_{\Delta}$ is the angle
between the in-plane component 
$\boldsymbol{\Delta}^{\perp} = (m_{1}+m_{2})(\boldsymbol{S}_{2}^{\perp}/m_{2}-\boldsymbol{S}_{1}^{\perp}/m_{1})$
and the infall direction at merger. The coefficients are obtained numerically
and are $H = 6.9\times10^{3}$, $A_{m} = 1.2\times10^{4}$, $B_{m} = -0.93$,
 $V_{1,1} = 3677.76$ km/s, $V_{A}=2481.21$ km/s, $V_{B}=1792.45$ km/s, and $V_{C}=1506.52$ km/s
\citep{lousto2012}. This fitting formula is similar to those obtained by previous authors,
e.g. \citet{campanelli2007a, baker2008, vanmeter2010}.
We adopt a value of $\xi = 145^{\circ}$ based on the
numerical results for quasi-circular merger configurations from \citet{lousto2008}.
The angle $\phi_{1}$ depends on the mass ratio and initial separation of the 
merging BHs, and can only be computed in numerical relativistic simulations; since
we are only interested in the statistical distribution of kick velocities, we take $\phi_{1} = 0$ 
and define $\phi_{\Delta}$ with respect to a fixed arbitrary direction, as 
recommended in \citet{lousto2010}. 

We assume randomly distributed spin magnitudes and spin
directions with respect to the orbital angular momentum as described in 
\S\ref{sec:merging}. The recoil velocity is highly sensitive to the 
orientation of the spins relative to the angular momentum. 

To calculate the trajectory of the kicked SMBH, we follow a similar 
prescription as \citet{madau2004}: we assume the density profile
of the galaxy is a truncated isothermal sphere, with a core radius equal to the radius of gravitational
influence of the post-merger SMBH, $R_{\mathrm{BH}} \approx GM_{\mathrm{BH}}/\sigma^{2}$,
so that $\rho(r) = \sigma^{2}/[2\pi G(r^{2}+R^{2}_{\mathrm{BH}})]$. 
The velocity dispersion $\sigma$ is obtained from Equation \ref{eqn1}.
If the $|\boldsymbol{V}_{\mathrm{recoil}}|$ from gravitational wave radiation is found
to be larger than the isothermal sphere escape speed $2\sigma (\ln (R_{e}/R_{\mathrm{BH}}))^{1/2}$,
where the effective radius $R_{e}$ is obtained from Equation \ref{eqn2}, we assume the black hole is ejected
from the galaxy. If the kick is insufficient to eject the black hole,
we calculate the time for the displaced SMBH to return to the center of the galaxy
via dynamical friction. 

We approximate the orbits of the kicked BHs
as purely radial. We solve numerically for the radial position of the SMBH as a function of time on a radial orbit 
using the equation of dynamical friction
\begin{flalign}
\frac{d^{2}\boldsymbol{r}}{dt^{2}} &= -\frac{GM(r)}{r^{2}}\boldsymbol{\hat{r}} \notag \\
&  -\frac{4\pi G^{2} \rho M_{\textrm{cusp}}\ln \Lambda}{v^{2}}\left(\textrm{erf}(x)-
\frac{2x}{\sqrt{\pi}}e^{-x^{2}}\right)\boldsymbol{\hat{v}},
\end{flalign}
where $M(r)$ is the mass within radius $r$, $x = v/\sqrt{2}\sigma$, and 
the Coulomb logarithm $\ln \Lambda$ is taken to be equal to 1, for reasons
described in \citet{madau2004} and \citet{maoz1993}. We take $M_{\mathrm{cusp}} =
2M_{\mathrm{BH}}$ for $V_{\mathrm{recoil}} < \sigma$ and $M_{\mathrm{cusp}} =
M_{\mathrm{BH}}$ otherwise. It should be noted that the 
dynamical friction times for radial orbits are highly dependent on the assumed
central density (or equivalently, core radius), and so are only a rough approximation.

As described in \citet{madau2004}, a kicked
BH would likely gain some angular momentum traveling through the galaxy so
that its orbit was no longer purely radial, which would lengthen
the time for it to return to the center. To test the impact of this,
we also calculated the dynamical friction time of all recoiling BHs using
Equation \ref{bt} with $f_{e} = 0.5$ and an initial radius of $R_{\mathrm{BH}}\exp
\left[(V_{\mathrm{recoil}}/2\sigma)^{2})\right]/\sqrt{e}$,
 and found the effect on our results to be negligible.

\section{Discussion}
\label{s41}

\subsection{Previous work}
\label{s1}

The evolution of supermassive black holes through cosmic time has been modeled using different
approaches to tracking the dark matter halos they reside in, including Monte Carlo merger trees
(e.g., \citealt{hk2000, haiman2009, nata2011}), Press-Schechter theory (e.g., \citealt{y2007}), and cosmological 
simulations (e.g., \citealt{dim2003, sijacki2009}). A more complete review of the early models of the 
growth of SMBHs is given in \citet{n2004}. 

Some works have focused purely on the feedback between the AGN and its galaxy without
consideration of mergers. One example is the semi-analytic model of \citet{granato2004}. 
In this model, star formation causes accretion onto the central
black hole via radiation drag; the black hole consequently emits radiation and kinetic outflows which
expel gas and quench star formation. Star formation also results in supernovae,
which similarly expel the gas. The feedback from supernovae and AGN results in galaxy
downsizing. The galaxy first goes through a phase 
as a dust-shrouded submillimeter galaxy (SMG)
involving rapid star formation and accretion onto the black hole,
which then induces feedback that causes the galaxy to become red and early-type. 
The model has been able to match observed properties of galaxies,
such as the masses and accretion rates of SMGs \citep{granato2006},
and the local early-type galaxy scaling relations \citep{cirasuolo2005} as well
as their size evolution \citep{fan2008}. 
It is also able to reproduce the mass functions \citep{granato2004}
and hard X-ray and optical luminosity functions \citep{lapi2006} of quasars.

Most studies have focused on understanding
black hole growth in individual galaxies residing in average over-density environments, typically the field. There have been
few studies of black hole mass assembly in extremely over-dense cluster and under-dense void environments.  Using 
dark matter merger trees derived from Press-Schechter theory, \citet{y2007} examined the expected growth of black
holes due to mergers in a large cluster with halo mass $10^{15} h^{-1} M_{\odot}$ at late times. They
found that most SMBHs in the cluster with masses $\gtrsim 10^{7.5} M_{\odot}$ underwent mergers,
with the most massive SMBHs increasing their mass by approximately a factor of 2 since $z = 2$. Although
the central galaxy generally contained the most massive SMBH at $z = 0$, \citet{y2007} found that
for some cluster assembly histories, the most massive SMBH may be hosted by a satellite galaxy.

Current simulations do not agree on the evolution of BH-galaxy scaling relations with redshift. Some
find that SMBHs at higher redshift should be more massive than if they followed the
$z = 0$ $M_{\bullet}-\sigma$ relation \citep{hop2009, dubois2012}, while others find that they should be less massive
\citep{malbon2007, matteo2008}. Some observational studies of 
broad line AGN samples at redshifts up to $z \sim 4$ find 
that these SMBHs are more massive at fixed
velocity dispersion than those at $z = 0$ \citep{mb2009, woo2008, greene2010},
whereas others report no significant evolution with redshift \citep{shields2003, gaskell2009, salviander2013}.
Similar studies have been done on the $M_{\bullet}-M_{*, \mathrm{bulge}}$
and $M_{\bullet}-M_{*, \mathrm{host}}$ relationships,
and have argued for evidence supporting an increase in black hole mass at fixed host mass with
increasing redshift \citep{merloni2010, b2011}. However, it has recently
been claimed that these observed trends in the BH-bulge scaling relations
may be the result of selection effects in observations of broad line AGN,
and that the observations are consistent with a lack of evolution \citep{schulze2013}.
Other works have investigated the evolution in the $M_{\bullet}-\sigma$ and
$M_{\bullet}-M_{*, \mathrm{bulge}}$ relations by combining observationally-derived bulge
mass and velocity dispersion functions at different redshifts with predicted
SMBH mass functions at the same redshifts. The latter are derived from 
observed AGN luminosity functions using the Soltan argument and some observationally-motivated
assumptions about SMBH merger rates. These studies do not have the same
observational biases as those using broad line AGN samples. They find
no evolution in the  $M_{\bullet}-\sigma$ relation \citep{shankar2009b, zhang2012},
but a larger $M_{\bullet}$ at fixed $M_{*, \mathrm{bulge}}$
with increasing redshift \citep{zhang2012}.

Because we fix the relations of seed mass to host bulge mass
and accreted mass to bulge star formation to be the same at all redshifts, we essentially
force the $M_{\bullet}-M_{*, \mathrm{bulge}}$ relation to be constant with redshift. Accounting for ejected and 
orbiting BHs can cause some evolution in the relation by lowering the central BH mass
relative to the host bulge mass, but, as will be described in \S\ref{s4},
the change in mass from these effects is on average fairly small. Also, because we take a fixed 
relation between galaxy stellar mass and velocity dispersion with
a redshift dependence $\sigma \propto (1+z)^{0.47}$, the $M_{\bullet} - \sigma$
relation as a function of redshift is also completely determined by our assumptions, which 
result in a declining BH mass at fixed $\sigma$ with increasing redshift. Thus our model
cannot reproduce evolution in the BH-galaxy scaling relations. The effect of this on our
results is similar to that of assuming an unevolving galaxy bulge mass fraction at fixed stellar mass,
as described in \S \ref{sec:evolution}; however, while assuming a constant bulge mass fraction
at all times causes SMBHs to be too massive at early times, assuming a constant
$M_{\bullet} - M_{\mathrm{bulge}}$ relation causes them to be less massive at early times.
 
Complementary constraints on BH growth models can be derived from X-ray luminosity functions of AGN. Since X-rays 
can escape even the most obscured Compton-thick galactic nuclear sources, they offer a unique 
probe of actively growing BHs \citep{salvaterra2006}. Several studies that have attempted to explain the origin
of the cosmic X-ray background have also provided insights into both the obscured and unobscured accreting
populations of BHs over cosmic time (c.f. models by \citealt{gilli2007, treister2009}).  Relevant to our study are 
the constraints obtained by \citet{volonteri2006}  on the accretion history of SMBHs at $z < 3$ using 
observations of the faint X-ray background combined with optical and hard X-ray luminosity functions. They
found that a model in which the Eddington ratio is a function of the AGN luminosity --- as suggested by
previous simulations --- fits the observational constraints somewhat better than models with
a constant Eddington ratio or an Eddington ratio decreasing with redshift.
However, other models have found that the Eddington ratio must depend on both AGN luminosity
and redshift in order to produce a high enough AGN fraction at low redshifts to match 
observations \citep{merloni2008, shankar2013}.

Similarly, \citet{natarajan2008} used the observed AGN X-ray luminosity function from $3 < z < 0$, combined with a simple
model of AGN accretion, to evolve seed black holes. They found that ultra-massive black holes (UMBHs)
with masses of $\sim 10^{10} M_{\odot}$ should exist at $z=0$, but that consistency with the observed present-day
SMBH mass function requires that there be an upper limit to the masses of SMBHs at about $10^{10} M_{\odot}$.
Such an upper limit could be the result of SMBH self-regulation processes, and would result in 
an evolving slope at the high end of the $M_{\bullet}-\sigma$ relation. 
They predict an abundance of UMBHs of $\sim 10^{-6} - 10^{-7} \, \mathrm{Mpc}^{-3}$, consistent
with extrapolating the $z=0$ SMBH mass function to high masses, and propose that UMBHs may
be found in central brightest cluster galaxies. Further work on the BHs hosted by central cluster galaxies (CCGs)
was done by \citet{volonteri2013}. SMBHs in such galaxies tend to be overmassive compared 
to the scaling relations for lower-mass galaxies, although the difference is more pronounced in
the $M_{\bullet}-\sigma$ relation than in the $M_{\bullet}-M_{\mathrm{bulge}}$ relation. They 
created semi-analytic models for the growth of
 SMBHs in CCGs and found that these trends may be the result of a larger number of dry mergers 
contributing to the mass of CCGs, because dry mergers increase a galaxy's mass, luminosity, and radius
more than its velocity dispersion. Minor mergers can in fact decrease the velocity dispersion
of the merged galaxy (\citealt{hilz2012}, Figure 9).

As for the population of ``wandering'' black holes, the buildup of populations of orbiting and ejected black holes as a result of galaxy
mergers and gravitational wave recoils has been examined most recently by \citet{rashkov2013}. Although their approach bears 
similarities to ours, they studied the evolution of intermediate-mass black holes (IMBHs), which
are thought to be the ancestral seeds of the supermassive black holes currently found in the centers
of galaxies. They populated the N-body \textit{Via Lactea II} cosmological simulation of a Milky Way-size
halo \citep{diemand2008} with seed IMBHs, which they then allowed to evolve via mergers and 
gravitational wave recoils. They found that even when assuming ``maximal'' numbers
of BHs would escape the galaxy due to gravitational wave kicks, a sizable population of leftover IMBHs 
should be orbiting in the halo of a galaxy with the mass of the Milky Way. We focus on slightly different
mass scales in this work and examine the central SMBHs and the SMBH wanderers in a 
typical cluster rather than a galaxy scale halo.

\subsection{Results}
\label{s4}

In this section, we present the results of following the SMBH evolution in both the cluster and void
environments. We also present some results for the ``global average''
combination of the cluster and void (\S \ref{scaling}), intended to be 
similar to the universe on average.

\begin{figure}[tbp]
  \begin{center}
    \includegraphics[width=\columnwidth]{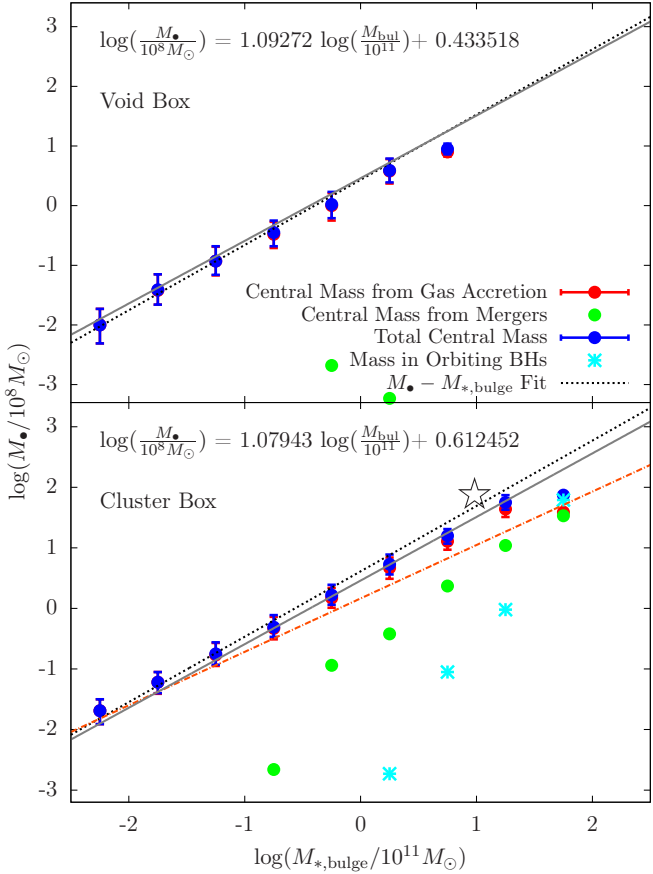} 
\caption[$M_{\bullet}-M_{*, \mathrm{bulge}}$ relation at $z=0$.]{$M_{\bullet}-M_{*, \mathrm{bulge}}$ relation at $z=0$.
The points are median masses and the error bars represent one quartile.
 We plot the median BH mass per galaxy in four categories: the total mass
of the central SMBH ($M_{\bullet}$), the amount of mass in the central SMBH that was accreted
from galactic gas, the amount of central SMBH mass from mergers with less massive SMBHs, and the
total amount of mass found in SMBHs that are orbiting in the galaxy. The black dashed line is
a fit to $M_{\bullet}-M_{*, \mathrm{bulge}}$ for \textit{all} the galaxies in each simulation box (not only
the displayed median points), and the resulting relation is displayed on the plot. 

The gray solid line is the observationally derived  $M_{\bullet}-M_{\mathrm{bulge}}$ relation from
\citet{mcc2012}. The orange dot-dashed line is a fit to $M_{\bullet}-\sigma$ data from the Virgo cluster
from \citet{ferrarese2006}, converted to $M_{\bullet}-M_{*}$ using Equation \ref{eqn1}. The star
symbol represents the position of M87 and its SMBH \citep{gebhardt2009, forte2012} marked as a 
reference.}
    \label{graph:graph1}
  \end{center}
\end{figure}

Figure \ref{graph:graph1} shows the median black hole mass in different categories versus the
bulge stellar mass in the void and cluster boxes. As expected
given our assumptions for SMBH growth, we reproduce the empirically observed
trend of central black hole mass with bulge stellar mass. Shown for comparison are the $M_{\bullet}-M_{\mathrm{bulge}}$
relation from \citet{mcc2012}, and the $M_{\bullet}-M_{*}$ relation from the Virgo cluster \citep{ferrarese2006}.
We find that in the cluster environment for the
most massive black holes a larger fraction of their mass growth occurs due to direct black hole
mergers compared to lower mass black holes. In the void box, where there are on average fewer
mergers, the median contribution of direct mergers to the mass inventory is less 
than $10^{6} M_{\odot}$ per galaxy at all bulge masses, whereas in the cluster it is $>10^{6} M_{\odot}$ for 
$M_{*, \mathrm{bulge}} > 3\times 10^{10} M_{\odot}$.
 
For the same reason, the amount of mass in BHs orbiting in galaxies in the void box is negligible,
as opposed to the cluster box, where the orbiting mass is substantial for galaxies with
$M_{*, \mathrm{bulge}} \gtrsim 10^{11} M_{\odot}$. 
In the cluster box, $\sim 2$ orbiting SMBHs with $M_{\bullet} \gtrsim 3\times10^{3} M_{\odot}$ 
are expected for each galaxy with mass $M_{*,\mathrm{bulge}} \approx 10^{11} M_{\odot}$  
($M_{*} \approx 1.3\times10^{11} M_{\odot}$), with the number increasing for larger masses. 
For the void box, $\sim 0.3$ orbiting BHs with $M_{\bullet} \gtrsim 3\times10^{3} M_{\odot}$ 
are found in the average galaxy with $M_{*,\mathrm{bulge}} \approx 10^{11} M_{\odot}$. 
It should be noted that the number of orbiting BHs per galaxy depends highly on
the minimum SMBH mass considered, since the dynamical friction time for a SMBH to reach
the center of a galaxy is inversely proportional to its mass (Equation \ref{bt}).

\begin{figure}[tbp]
  \begin{center}
    \includegraphics[width=\columnwidth]{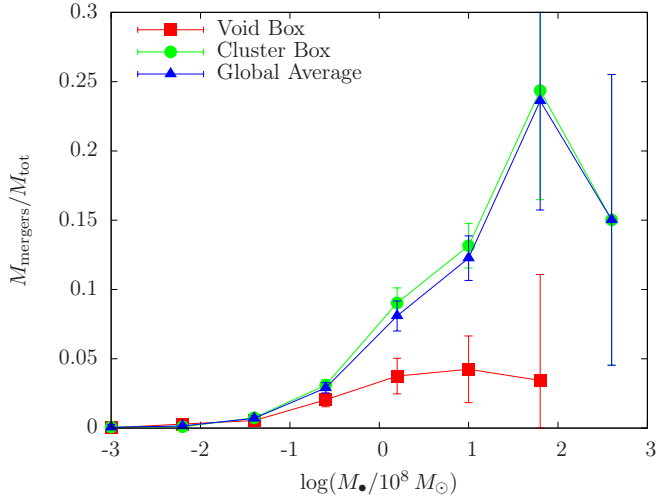}
 \caption{The number-weighted mean fraction of total central SMBH mass gained via BH-BH
mergers as a function of SMBH mass at $z=0$.
	Error bars are $1\sigma$.
Values are shown for BHs in the void and cluster boxes as well as the global
average combination of the two.}
    \label{graph:graph2}
  \end{center}
\end{figure}

The difference between the fraction of BH mass contributed by mergers in the void and cluster
boxes is shown more clearly in Figure \ref{graph:graph2}. 
The number-weighted mean mass fraction from mergers is essentially zero for
SMBHs with $M_{\bullet} < 10^{6} M_{\odot}$ in both boxes and increases for larger 
masses. In the cluster, the fraction reaches a maximum of $24\%$ 
for $M_{\bullet} \approx 10^{10} M_{\odot}$.
In the void, the fraction of mass from mergers reaches about $5\%$ for the most massive BHs. 
The fraction of central black hole mass from mergers can also be examined as a function of the 
host bulge mass. The contribution from mergers is negligible for host masses
$M_{*,\mathrm{bulge}} \lesssim 10^{9.5} M_{\odot}$ in both the void and cluster boxes.
It increases with host mass such that in the void the fraction of 
BH mass from mergers is $2\%$ at $M_{*,\mathrm{bulge}} = 10^{10} M_{\odot}$,
$4\%$ at $10^{11} M_{\odot}$, and $19\%$ for
the most massive galaxies in the void, which have masses about $10^{12} M_{\odot}$.
In the cluster box, the fraction from
mergers is $4\%$ at $M_{*,\mathrm{bulge}} = 10^{10} M_{\odot}$,
$11\%$ at $10^{11} M_{\odot}$, and $23\%$ at $10^{12} M_{\odot}$, reaching
a maximum of $28\%$ for the most massive galaxies with $\approx 10^{13} M_{\odot}$.
The latter is close to the fraction of galaxy mass attained via mergers 
by the most massive galaxies in our cluster box, which is $\sim 40 \pm 15 \%$ \citep{lack2012},
a value that is significantly lower than that found by some previous simulations \citep{oser2010}.
The fact that the BH mass fraction from mergers is slightly smaller 
than that for galaxies is expected due to a combination of two factors.
The first is that the mass of our SMBHs is roughly
proportional to their host bulge stellar masses,
which increase as a fraction of total stellar mass for larger galaxies.
Thus the stellar mass ratio of a smaller galaxy
to a larger one will tend to be larger
than the ratio of the masses of their respective SMBHs, resulting
in a larger contribution from mergers for galaxies
than for the SMBHs that they host. The other effect is that
some fraction of the BHs from merged galaxies are still orbiting at $z=0$.
 The number-weighted mean mass contribution from mergers 
for all SMBHs is $0.9 \pm 0.1\%$ in the void box and 
$2.4 \pm 0.1\%$ in the cluster box. 

Between redshifts $z = 4$ and $z=0$, the mean mass-weighted merger ratio 
for SMBH mergers in both the void and cluster is $\sim 1:5$. Our
results can be compared to the median mass-weighted
merger ratio of galaxy bulges from \citet{hopkins2010},
who studied galaxy bulges with masses $10^{9} M_{\odot} \lesssim M_{*, \mathrm{bulge}} \lesssim 10^{12} M_{\odot}$
at $z = 0$ using semi-empirical models. Our results are consistent
with their values for the median merger ratio, 
which span approximately $1:6$ to $1:2$ depending on 
the final bulge mass. We also compare 
with the mean mass-weighted merger ratio for individual galaxies found by \citet{oser2011},
who performed cosmological re-simulations of galaxies with masses $4.5 \times 10^{10} h^{-1} M_{\odot}
\lesssim M_{*} \lesssim 3.6 \times 10^{11} h^{-1} M_{\odot}$
between $z = 2$ and $z = 0$.
Our mean SMBH merger ratios are the same
for $2 > z > 0$ as for $4 > z > 0$, and agree with
the value of $\sim 1:5$ obtained by \citet{oser2011} for galaxies.
The agreement with \citet{hopkins2010} and \citet{oser2011} is likely a result of the fact that the mean mass-weighted galaxy merger
ratio in our simulations is approximately in agreement
with those of the referenced works \citep{lack2012}.
Since SMBH mergers are subsequent
to galaxy mergers and SMBHs and their hosts are connected by scaling relations, the similarity
of the results is not unexpected. 

\begin{figure}[tbp]
  \begin{center}
    \includegraphics[width=\columnwidth]{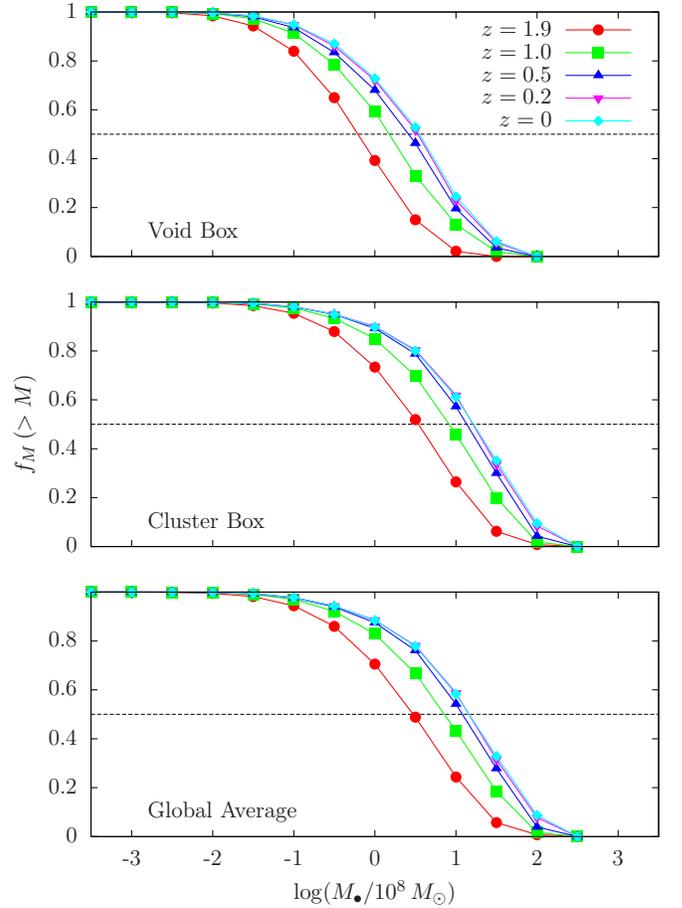} \caption{The fraction of the total central black hole
    mass contained in black holes with mass above some $M_{\bullet}$,
		shown at different redshifts.}
    \label{graph:graph6}
  \end{center}
\end{figure}

Figure \ref{graph:graph6} shows the distribution of central SMBH mass
in the form of the fraction of total black hole mass contained in BHs with mass $> M_{\bullet}$.
In both the cluster and void box, the total mass is overwhelmingly concentrated in the most
massive black holes. 
At low redshifts, half of the mass is found 
in SMBHs with $M_{\bullet} \gtrsim 10^{8.5} M_{\odot}$ in the void,
and in SMBHs with $M_{\bullet} \gtrsim 10^{9} M_{\odot}$ in the cluster, despite the fact that
such black holes constitute less than $5\%$ of the total population in both boxes.
The effects of BH-BH mergers can also be seen, as the fraction of mass contained in high mass
black holes increases over time up to $z=0.2$, but then stops as a result of more
low mass black holes being added to the population.

\begin{figure}[tbp]
  \begin{center}
    \includegraphics[width=\columnwidth]{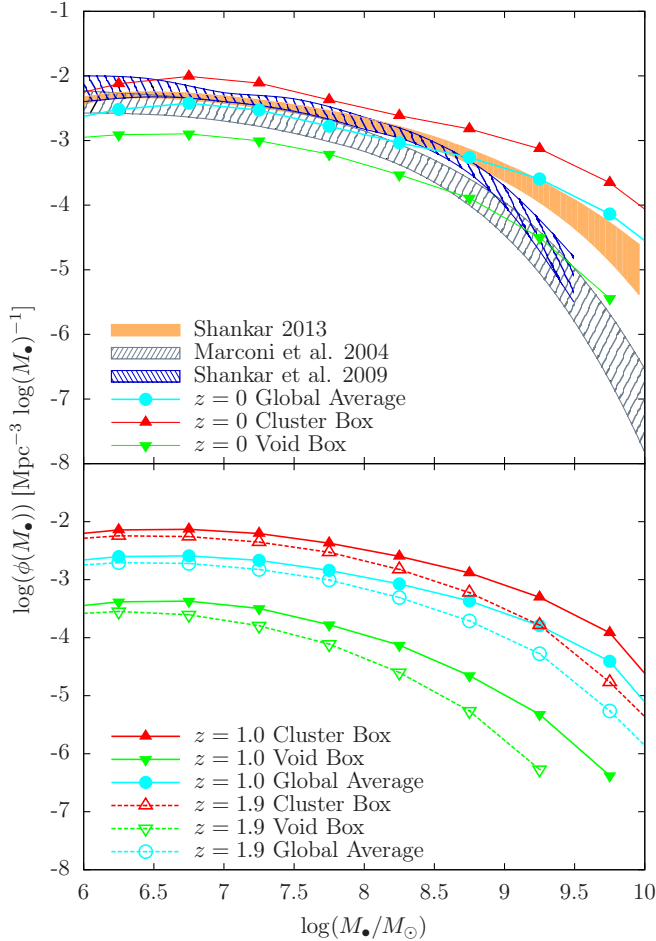} \caption[SMBH Mass Function]{\textit{Top Panel}: 
The mass function of central SMBHs at $z = 0$, for the cluster
box, void box, and global average. Also shown for
comparison are the observed SMBH mass functions from \citet{marconi2004}, \citet{shankar2009},
and \citet{shankar2013b}.

\textit{Bottom Panel}: Same as the top panel, but at $z = 1$ and $z = 1.9$.}
    \label{graph:graph19}
  \end{center}
\end{figure}

A different way of displaying the mass distribution of SMBHs can be
seen in Figure \ref{graph:graph19}, which shows the differential
mass function per logarithmic SMBH mass interval. Only central SMBHs are included. Unlike
Figure \ref{graph:graph6}, in which the fraction of SMBH mass in each bin
is normalized to the total SMBH mass in the box,
Figure \ref{graph:graph19} shows the number of SMBHs per unit volume in each mass
bin, which also displays the larger number density of SMBHs in the 
cluster box compared to the void box (which is due to the larger density of
galaxies in the cluster). In the top panel, we compare our global average SMBH mass function
at $z=0$ to that from observations \citep{marconi2004, shankar2009, shankar2013b}, and
also show the mass functions of the cluster and void boxes at $z = 0$.
The lower panel shows the cluster, void, and global average central SMBH mass
functions at $z = 1$ and $z = 1.9$. The number of 
SMBHs increases with time for all $M_{\bullet} > 10^{6} M_{\odot}$,
but especially for higher masses, in both the cluster and void.

One can see in the top panel of Figure \ref{graph:graph19} that our
 $z=0$ mass function is shallower than 
those of \citet{marconi2004} and \citet{shankar2009}, but is nearly
 in agreement with a more recently derived mass function
from \citet{shankar2013b}. As described in 
\S \ref{scaling} and \S \ref{sec:evolution},
we evolve our SMBHs and normalize our SMBH mass densities in the cluster and void 
under the assumption that the $M_{\bullet} - M_{\mathrm{bulge}}$ relation in \citet{mcc2012} is correct.
The mass functions of \citet{marconi2004} and \citet{shankar2009} are derived
using a different $M_{\bullet}-M_{\mathrm{bulge}}$ relation,
which results in a different shape and lower normalization than our mass function.
The mass function from \citet{shankar2013b} 
is instead derived assuming that all SMBHs follow the  $M_{\bullet} - \sigma$ relation of \citet{mcc2012}, 
similar to our assumption. Thus the greater similarity of our mass function 
to that derived by \citet{shankar2013b} is not surprising.

\begin{figure}[tbp]
  \begin{center}
    \includegraphics[width=\columnwidth]{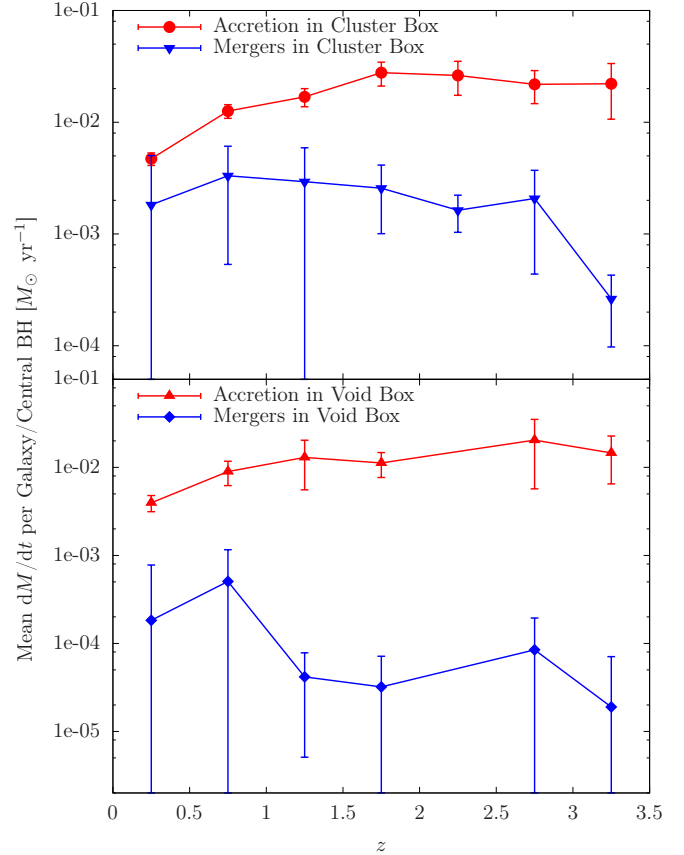} \caption{Number-weighted 
mean mass growth rate per central SMBH from accretion and mergers. 
The top and bottom panels show the cluster and void box, respectively. The red 
lines show the rate of mass increase due to gas accretion onto central black holes; the blue
lines show the rate of mass increase due to mergers with smaller black holes. The mean
mass of a SMBH at $z = 3.25$ is $4.5\times10^{6} M_{\odot}$ in the cluster and 
$1.4\times10^{6}M_{\odot}$ in the void, increasing to $9.3\times10^{7} M_{\odot}$ in the cluster and 
$2.5\times10^{7} M_{\odot}$ in the void at $z = 0.25$.}
    \label{graph:graph20}
  \end{center}
\end{figure}

The growth of SMBHs over time follows different trends in the void
and cluster boxes. Figure \ref{graph:graph20} shows the 
number-weighted mean rate of mass growth via accretion and mergers per
central SMBH as a function of redshift.
In both the cluster and void, accretion is the dominant source of  
growth for SMBHs at all redshifts. Nevertheless, the accretion
rate per BH decreases by a factor of $\sim 5$ and 
mergers become more significant in both boxes between $z = 4$ and $z = 0$. 
While the mean accretion rate per SMBH is only slightly higher in the cluster than in the void,
the mass growth rate from mergers is $\gtrsim 10$ times higher
in the cluster at all redshifts.

Because mergers become more significant over time, in both the void and cluster
 the total black hole mass from mergers is added slightly later on average than that from 
accretion. In the void, half of the total mass from accretion at $z = 0$ has been added by $z \approx 0.6$,
whereas for mergers half the mass has been added by $z \approx 0.5$. In the cluster,
both half the merged and accreted mass are added earlier than in the void,
and the difference between the two is larger: half the accreted mass is added 
by $z \approx 1.0$, while half the merged mass is added after $z \approx 0.7$.
These SMBH growth trends parallel those of galaxies.
Galactic mass assembly in the cluster tends to happen earlier than in the void, and in both the cluster and void
mass buildup by accretion peaks at earlier times than by mergers
\citep{lack2012}.

\begin{figure}[tbp]
  \begin{center}
    \includegraphics[width=\columnwidth]{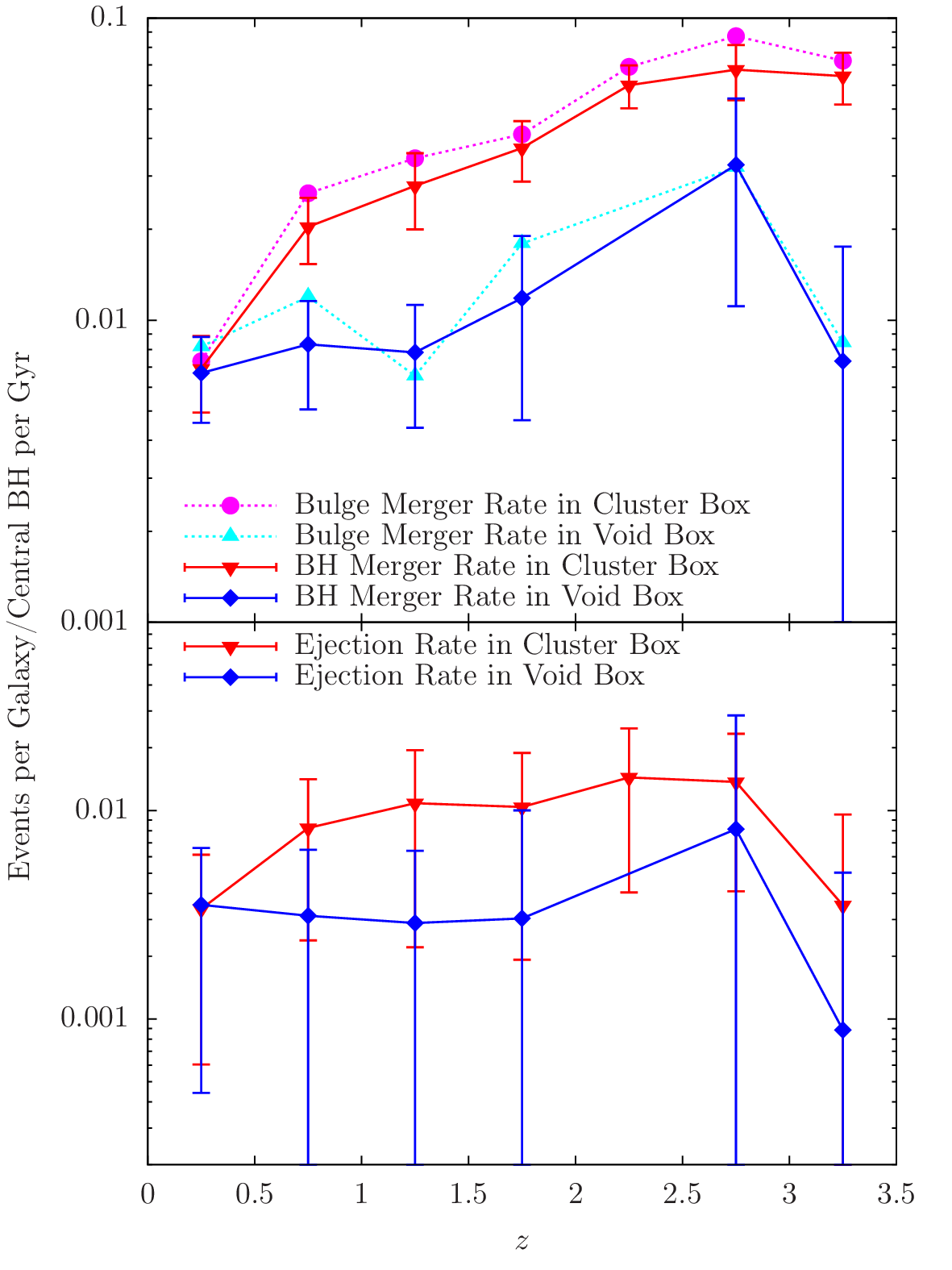} \caption[Merger and Ejection Rates]{\textit{Top Panel:} Merger rates 
of galaxy bulges and SMBHs, expressed as
the number of mergers per galaxy or central BH, 
weighted by merger ratio (e.g., a merger with ratio 1:2 is counted as 1/2), per Gyr.
Dotted lines represent the merger rate for bulges and solid lines represent the merger
rate for black holes.

\textit{Bottom Panel:}
The ejection rate of black holes from their host galaxies, expressed as the
number of ejections per Gyr.}
    \label{graph:graph21}
  \end{center}
\end{figure}

The connection between galaxy growth via galaxy mergers and 
BH growth via BH-BH mergers can also be seen in 
the top panel of Figure \ref{graph:graph21}. This panel shows the 
number of mergers, weighted by the merger mass ratio (of BHs or bulges), 
per central BH per Gyr. The merger history of the BH population
approximately follows that of the bulge population --- which is to be expected,
as BH mergers are subsequent to galaxy mergers, although sometimes
delayed by dynamical friction. One can also see in this panel that in
the cluster on average a central BH is predicted to
experience a mass increase due to mergers equivalent to a $\sim 1:10$ merger
per Gyr at $z \sim 3$, declining to the equivalent of a $\sim 1:100$ merger
per Gyr at $z \sim 0$. In the void, the mass gain rate from mergers is lower than
that in the cluster but declines with time more slowly, becoming
the same as for the cluster at $z \sim 0$. We predict a mass gain
in the void per average central BH equivalent to a $\sim 1:50$ merger per Gyr at $z \sim 3$
and a $\sim 1:100$ merger per Gyr at $z \sim 0$.
There is a factor of $\sim 3$ difference between the void and cluster in the fraction of
growth due to mergers per central galaxy for $z \gtrsim 0.5$; this is smaller than the factor of $\gtrsim 10$ 
seen in Figure \ref{graph:graph20} for $d M/ dt$ from mergers because SMBHs in the void
are on average less massive than those in the cluster.

The bottom panel of Figure \ref{graph:graph21} shows a related quantity, the fraction
of central BHs ejected from their host galaxies per Gyr as a function of redshift.
This follows the trend of the BH merger rate closely, since mergers cause 
ejections via gravitational wave recoil. The scatter is very large
in both the cluster and void, but the fraction of BHs ejected per Gyr
is fairly constant over time: in the cluster it is $\sim 0.01$, and in the void, $\sim0.003$.
 
The velocity of the BH recoil is sensitively dependent on the
angles that the spins of the merging BHs make with the
angular momentum of their orbit. When these spins are aligned, 
the kick velocity cannot exceed 200 km s$^{-1}$ \citep{bogdanovic2007}. 
However, even though we assume an isotropic distribution of spin vectors,
the fraction of ejected BHs is relatively small, as can be inferred from Figure \ref{graph:graph21} and 
will be described in further detail in \S\ref{soltan}. As a result,
even a complete lack of ejections will not have a large effect on the remaining
results. If we assume that all merging BHs are aligned to within $\sim 10^{\circ}$ of the orbital
angular momentum, as expected for gas-rich mergers in cold disks \citep{dotti2009}, 
generally no ejections occur in our realizations. Despite the smaller kicks out of 
the center of the galaxy, the number and mass in orbiting BHs is
not significantly affected. Relaxing the assumption to take all BHs to be aligned within
a larger angle of $\sim 30^{\circ}$, as in 
gas-rich mergers in warm disks \citep{dotti2009}, we find that the fraction
of BHs and BH mass ejected is roughly half that when assuming an isotropic
distribution of spins. The population of orbiting BHs is unaffected.
Thus, our assumptions about the directions of BH spins do not appear to affect our results
significantly.

\begin{figure}[tbp]
  \begin{center}
    \includegraphics[width=\columnwidth]{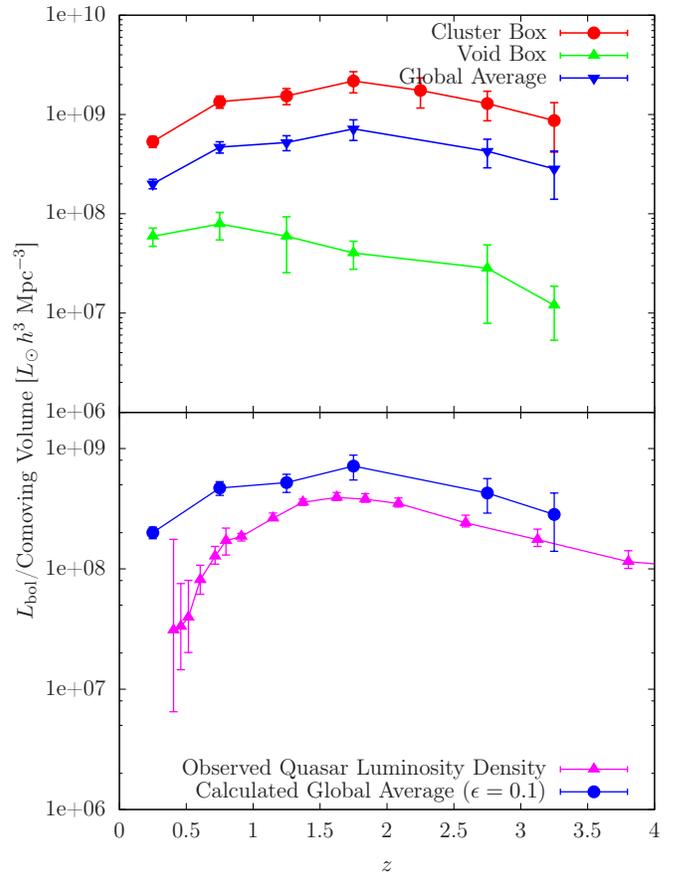} \caption[SMBH Bolometric Luminosity]
{\textit{Top Panel:} Estimated SMBH bolometric luminosity
    $E=0.1\dot{M}_{acc}c^2$ per unit comoving volume due to gas accretion as a function of $z$ for the void
and cluster boxes and their weighted global average.

\textit{Bottom Panel:} In magenta is the
observed QSO luminosity density from \citet{hopkins2007}. The blue solid
line is our calculation for the global average bolometric luminosity using $\epsilon = 0.1$,
 also shown in the top panel. }
\label{graph:graph4}
  \end{center}
\end{figure}

Figure \ref{graph:graph4} shows the expected bolometric luminosity density of SMBHs due to gas accretion
as a function of redshift for the cluster and void boxes as well as the global average.
We assume a mass to energy conversion efficiency of $\epsilon = 0.1$,
so that $L_{\mathrm{bol}} = 0.1\dot{M}_{\mathrm{acc}}c^{2}$. The
results can be scaled for other assumed values of the efficiency. 
In the lower panel of Figure \ref{graph:graph4}, we compare our estimated global average luminosity density to 
the observed  QSO luminosity density from \citet{hopkins2007}. 
The luminosity density we calculate is higher than
that of \citet{hopkins2007}. The mass to energy conversion efficiency we use
is consistent with that found in \citet{yutrem2002}, but their calculation
assumed a local black hole mass density of $\rho_{\bullet} = 2.5\times10^{5} M_{\odot}$ Mpc$^{-3}$,
whereas we use $9.1\times10^{5} M_{\odot}$ Mpc$^{-3}$, which would imply a lower value of $\epsilon$.
Thus the fact that our calculated luminosity density is higher than that observed
is not surprising.

Because our adopted $\rho_{\bullet}$ implies a lower radiative efficiency from the
Soltan argument, it is more interesting to look at the difference in shape of our SMBH luminosity
density with redshift as opposed to normalization. The global average luminosity output
we find is roughly constant with redshift for $z > 0.5$, decreasing for the most recent times.
However, the downturn we find at low redshift is much smaller than that seen in
 observations. This may be because 
the cosmological simulation we use is known to especially overproduce stars at low $z$
\citep{lack2012}, which would cause the BH accretion rate to remain too high as well.
Our scaling of the void to cluster is also quite simplistic, taking the void 
to cluster ratio to be constant with redshift. If this is not the case, the
shape of our predicted trend would also change.

\begin{figure}[tbp]
  \begin{center}
    \includegraphics[width=\columnwidth]{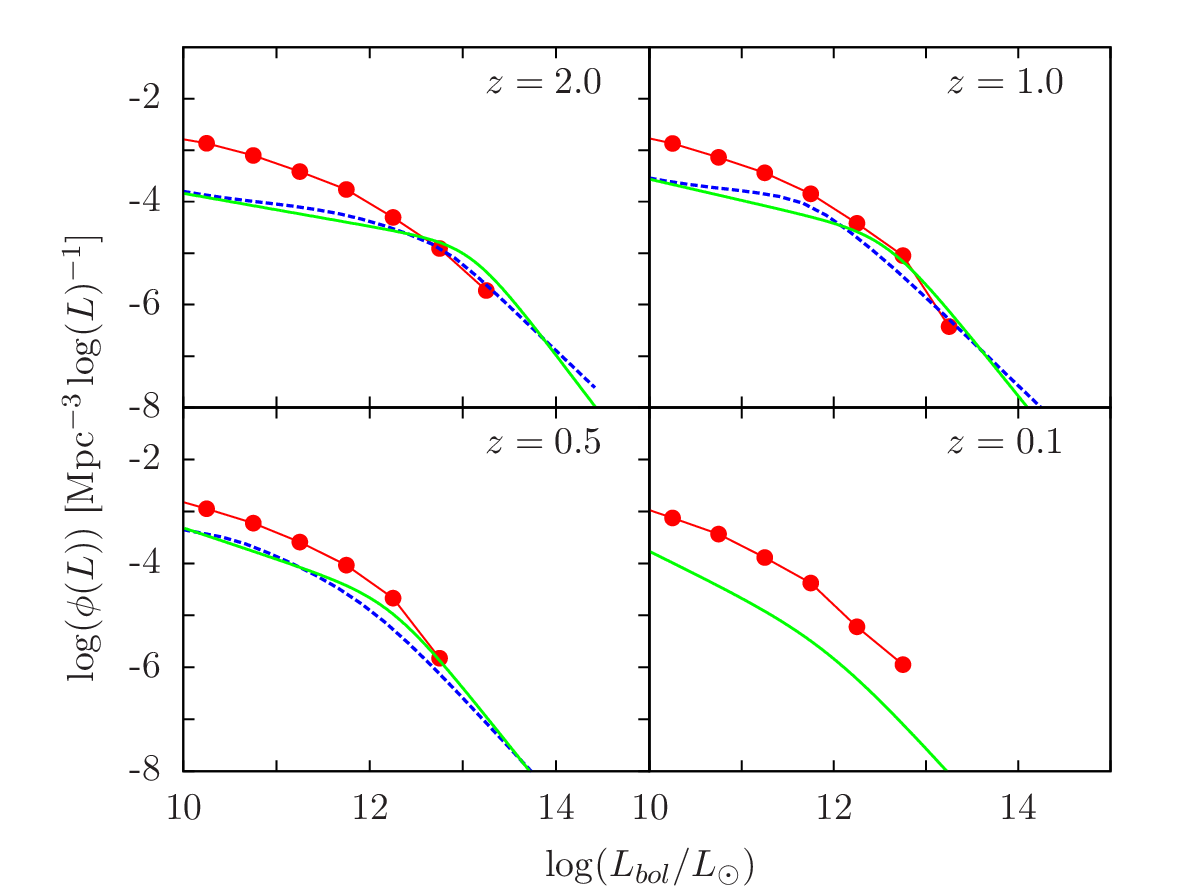}\caption{Global average instantaneous bolometric SMBH 
luminosity functions at $z=$ 2.0, 1.0, 0.5 and 0.1 (red points) compared with SMBH bolometric luminosity
functions derived from observations from \citet{hopkins2007} (green lines) and
\citet{shankar2009} (blue dashed lines; for $z \ge 0.5$ only).}
    \label{graph:graph91}
  \end{center}
\end{figure}

Figure \ref{graph:graph91} shows our global average instantaneous
bolometric SMBH luminosity function (LF) at redshifts
2.0, 1.0, 0.5, and 0.1. For comparison are shown model fits to observational data
from \citet{hopkins2007}. 
In Figure \ref{graph:graph4} it can be seen that the total luminosity output
we predict from QSOs is higher than that found by \citet{hopkins2007} and
\citet{shankar2009}, especially at $z < 0.5$.
Figure \ref{graph:graph91} shows the instantaneous LF as opposed to the mean
total luminosity in broader redshift bins shown in Figure \ref{graph:graph4},
but the integrated total luminosity we predict is still larger
than that obtained from the LFs of \citet{hopkins2007} and \citet{shankar2009}, 
especially at $z = 0.1$. 

Also, our LF is not of the same shape as that observed; it is 
skewed towards low luminosities. This is because our simple assumptions
for the SMBH growth, combined with the limitations of our simulation, are unable to
produce very high luminosity QSOs and assign 
luminosities that are too low to some fraction of SMBHs instead.
 We assume that the instantaneous SMBH accretion rate 
is proportional to the instantaneous star formation rate of its host,
and that there is a constant proportionality between
 the instantaneous SMBH accretion rate and the instantaneous luminosity, such that 
the latter is always $\epsilon = 0.1$ times the former.
Our model includes no detailed assumptions about accretion physics onto the
SMBH that could cause large variability in the SMBH luminosity.
Furthermore, our simulation models the star formation rate as exponentially declining 
for each star particle \citep{cenostriker1992}, 
so it does not capture the observed ``bursty'' nature of star formation, which would increase
the number of high-luminosity QSOs during bursts in the simple scheme described above.
Thus our model has no mechanism for producing the brightest QSOs. 
If we included the mechanisms above, some of the SMBHs in our model
would be assigned higher luminosities, which would increase the number of high-luminosity QSOs relative
to low-luminosity ones and decrease the discrepancy in LF shape
between our results and observations.

\begin{figure}[tbp]
  \begin{center}
    \includegraphics[width=\columnwidth]{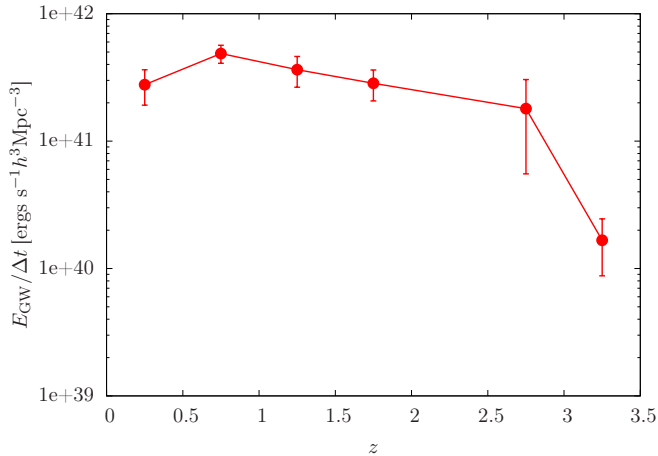}\caption{The global average gravitational wave
    luminosity density resulting from BH-BH mergers.}
    \label{graph:graph10}
  \end{center}
\end{figure}

We also calculate the expected total energy density emitted in gravitational wave radiation
as a result of BH-BH mergers. The results are shown in Figure \ref{graph:graph10} for the global average.
The luminosity follows a general increase with time, with the increase being
significantly steeper for $z \gtrsim 2.5$. The gravitational wave luminosity, while
very similar as a function of redshift to the rate of mass increase due to mergers, is not directly proportional, since
the energy in emitted gravitational wave radiation is also dependent on the mass ratio of the merging
black holes (see Equations \ref{gwfirst}-\ref{gwlast}). 

Further, we calculate the gravitational wave strain produced by our population of black holes using the method
described in \citet{sesana2012}. For a population of merging black hole binaries,
where the black holes in each binary have masses $M_{1}$ and $M_{2}$ with $M_{1} > M_{2}$ and $q \equiv M_{2}/M_{1}$,
the characteristic amplitude of the gravitational wave signal $h_{c}$ is given by
\begin{equation}\label{h1}
\mkern-15mu h_{c}^{2}(f){=}\frac{4}{\pi f} \iiint \! \der z \der M_{1} \der q \frac{\der^{3} n}{\der z \der M_{1} \der q}
\frac{1}{1+z}\frac{\der E_{\mathrm{gw}}(\mathcal{M})}{\der \ln f_{r}}
\end{equation}
where the energy emitted per logarithmic frequency interval is
\begin{equation}\label{h2}
\frac{\der E_{\mathrm{gw}}}{\der \ln f_{r}} = \frac{\pi^{2/3}}{3}\mathcal{M}^{5/3}f_{r}^{2/3}
\end{equation}
where $\mathcal{M} = (M_{1}M_{2})^{3/5}/(M_{1}+M_{2})^{1/5}$ is the chirp mass of the binary
and $f_{r} = (1+z)f$ is the rest frame frequency of the gravitational radiation.
The amplitude $A$ is defined by 
\begin{equation}
h_{c}(f) = A\left(\frac{f}{\mathrm{yr}^{-1}}\right)^{-2/3}.
\end{equation}
We find a global average strain amplitude
of $\log A = -14.7 \pm 0.1$ ($1\sigma$ error), or
$A = 2.0\times10^{-15}$. This is below the current observational upper limit 
of $A = 6\times 10^{-15}$ found by \citet{vh2011}. Our result is 
consistent with the $1\sigma$ range found by 
\citet{sesana2013, sesana2012} for models using the $M_{\bullet}-M_{\mathrm{bulge}}$
relation of \citet{mcc2012} as we do. Our results are within the $2\sigma$ lower
limit of \citet{mcwilliams2012} but lower than their main result of $A = 4.1\times10^{-15}$.
This is probably because \citet{mcwilliams2012} computed the merger rate
assuming all late-time growth to be due to mergers, whereas in our model
late-time growth is still dominated by accretion (Figure \ref{graph:graph20}).
For a $f^{-2/3}$ gravitational wave spectrum from inspiraling 
binaries, this value for the strain could potentially be detected with
$\lesssim 8$ years of pulsar timing observations at current sensitivities,
or fewer with higher sensitivities (\citealt{sesana2013}, Figure 2).

It should be noted that our simulations suffer from overmerging of galaxies \citep{lack2012}.
Also, because the galaxies in our simulation
have higher stellar masses than the scaled-down ones we use
for our calculation, our galaxy merger rates are further overestimated.
This is apparent from the dependence on stellar mass of 
the dynamical friction time for a galaxy in a subhalo to merge 
with the central galaxy in an NFW dark matter halo; for this case
$t_{\mathrm{DF}} \propto (1+M_{*}/M_{\mathrm{DM}})^{-9}$ \citep{mcwilliams2012b}.
This may lead to an overestimation of the number of BH-BH mergers, which
would cause us to overestimate the gravitational wave luminosity
and strain. It would also increase the predicted number of orbiting
and ejected SMBHs; these results are described further below.

\subsubsection{Scatter in the $M_{\bullet}-M_{*, \mathrm{bulge}}$ relation with mass}

\begin{figure}[tbp]
  \begin{center}
    \includegraphics[width=\columnwidth]{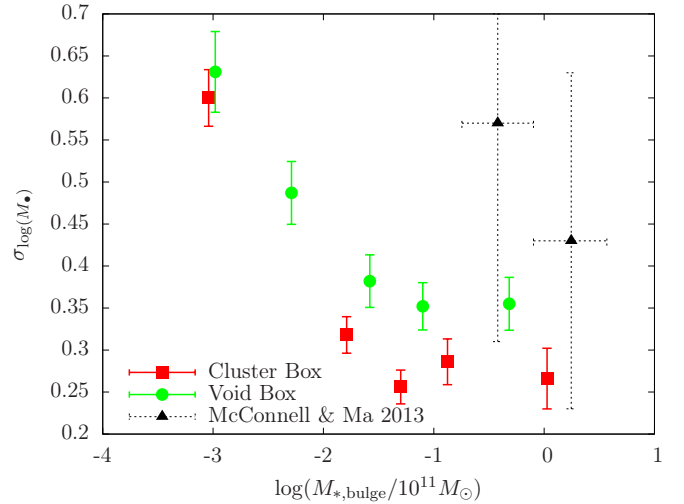} \caption{Scatter around the $M_{\bullet}-M_{*,\mathrm{bulge}}$
relation as a function of $M_{*,\mathrm{bulge}}$ for the cluster and void boxes at $z=0$. Points represent
	the scatter $\sigma$ around a separate $M_{\bullet}-M_{*,\mathrm{bulge}}$ relation fit in each bin. 
Error bars are one standard deviation of $\sigma$. Also plotted as black triangles with dotted error bars 
are the values of the intrinsic scatter from Bayesian fits to the $M_{\bullet}-M_{\mathrm{bulge}}$ relation 
in separate mass bins from \citet{mcc2012}.
}
    \label{graph:graph3}
  \end{center}
\end{figure}

We divide the galaxies in the cluster and void boxes at $z=0$ into 
bins in bulge stellar mass, and fit separate
$M_{\bullet}-M_{*, \mathrm{bulge}}$ relations to each bin. We then calculate the 
scatter around the relation in each bin, $\sigma$, given by
\begin{equation}
\sigma^{2} = \frac{\sum\limits_{i} \left[\log_{10}(M_{\bullet , i}) - \alpha - \beta x_{i} \right]^{2}}{N_{dof}},
\end{equation} where $\alpha$ and $\beta$ are the coefficients of the fit. We plot the scatter for
each bin in Figure \ref{graph:graph3}.
The relation is found to be tighter for black holes with larger masses in both the void
and cluster box. The current sample of SMBHs with observationally measured masses is not
sufficiently large to determine whether a such a decrease is seen in the scatter. The most complete
sample to date is found in \citet{mcc2012}, who calculate the 
intrinsic scatter in the $M_{\bullet} - M_{\mathrm{bulge}}$ relation using several different methods.
In Figure \ref{graph:graph3} we plot their estimates
for the intrinsic scatter from Bayesian fits to separate mass bins, as this method is the most similar
to ours because it involves fitting the individual mass bins with separate $M_{\bullet}-M_{\mathrm{bulge}}$
relations. The values of the scatter we obtain are slightly in tension
with those found by \citet{mcc2012}, being significantly smaller than their
mean values for the intrinsic scatter but marginally consistent with their large error bars.

A decrease in scatter with increasing bulge or BH mass has been proposed 
as an expected result of galaxy and subsequent BH-BH mergers, which tighten the black hole scaling relations
due to the central limit theorem \citep{h2010}. However, in our model,
mergers contribute only a small amount to this decrease in scatter. The majority
of the decrease originates from the fact that more massive SMBHs tend to 
have had more episodes of accretion, which, because it is proportional to 
the star formation in the host galaxy bulge, also lowers the scatter in the 
$M_{\bullet}-M_{*,\mathrm{bulge}}$ relation by the central limit theorem. Because 
our merger trees have only a few discrete time slices at which we grow every existing SMBH
via accretion, the most massive SMBHs are also those
that were seeded the earliest. However, this is simply a limitation of our model,
and the fact that BHs which experience more 
star formation-related accretion episodes
will be both more massive and have tighter BH-galaxy scaling 
relations is true independent of the length of time a SMBH has existed.

\subsubsection{Corrections to the Soltan argument}
\label{soltan}

As described in \S \ref{s2}, the Soltan argument 
depends on our ability to accurately measure the mass density in SMBHs at the present
day. Any luminosity emitted via the accretion of mass onto an SMBH will
be included in the inventory of total SMBH luminosity produced over
all time. However, if some fraction of SMBHs end up orbiting or ejected by the present day,
some fraction of the mass accreted onto SMBHs 
will not be included in an inventory of local SMBH mass density, which only counts the SMBH mass in galactic centers. 
The fraction of SMBH luminosity produced by the accretion of such mass 
will be assumed to have come from 
the SMBHs that are found in the centers of galaxies, causing their mean radiative
efficiency $\epsilon$ to be overestimated.
In our model, we keep track of the fraction of mass in orbiting and ejected BHs that
must be a correction to the Soltan argument. We present these fractions 
below. We also give ranges for some of the results we 
obtain in multiple realizations of the model
to emphasize that some of these results can vary widely between our different realizations
and so should be taken as uncertain.

We find that at $z=0$, in the void, the fraction of all BHs that are orbiting in the outskirts
of their host galaxies is $3.2\pm0.7\%$, with a range between our different realizations of
$1.6\%$ to $5.0\%$. This includes all BHs with masses $M_{\bullet} \gtrsim 3\times10^{3} M_{\odot}$.
The fraction of the total SMBH mass in orbiting black holes is 
$0.11\pm0.06\%$, with a large range $0.007\%$ to $0.24\%$. For the cluster, the amount of orbiting mass is much
greater than in the void, as can already be seen in Figure \ref{graph:graph1}. In fact,
there is a very large population of orbiting SMBHs in the cluster, with $40.4\pm0.4\%$ 
of BHs at $z=0$ orbiting; however, most of these BHs are quite small
 (since less massive BHs have longer dynamical friction
times), and so the fraction of mass in orbit is $7.6\pm1.8\%$,
with range $4.3\%$ to $13.9\%$. It should be noted that these
orbiting black holes are highly concentrated in the most massive galaxies in our cluster
box --- as could be expected from the fact that a larger galaxy will have a longer dynamical friction
time. In fact, approximately one third of all orbiting
BHs in the cluster are found in the most massive galaxy. 
Combining the cluster and void, these values correspond 
to a global average fraction of orbiting BHs of 
$35.7\pm0.4\%$.
 The fraction of mass that is unaccounted
for by observing galaxy centers is $7.0\pm1.6\%$, 
with range $4.0\%$ to $12.7\%$.

We also predict the fraction of black holes that are completely ejected from their host galaxies due 
to gravitational wave recoil. At $z = 0$, we find that in the void $2.3\pm0.7\%$
of all black holes are not associated with any galaxy as a result of being ejected,
equivalent to a loss of $2.2\pm2.2\%$ of the total SMBH mass with very
large range $0.1\%$ to $9.9\%$. In the cluster $4.1\pm0.5\%$ of all BHs and 
$4.4\pm3.2\%$ (range $1.2\%$ to $13.9\%$) of all BH mass is ejected. It should be noted that
while the fraction of black holes ejected is nearly constant for all our runs of the model, 
the fraction of mass ejected can vary significantly from one to the other. 
The global average fraction of ejected BHs is 
$3.9\pm0.4\%$, and the fraction of mass ejected is 
$4.2\pm2.9\%$ with a range 
$1.2\%$ to $12.8\%$.
 
In combination, we predict that the total correction to the Soltan argument 
from both unaccounted-for orbiting and ejected BHs is within the range 
$5.9-20.5\%$, on average $11.2\pm3.4\%$.

\section{Conclusions}

We have used the results of a set of hydrodynamical cosmological simulations of a void and cluster
region in the universe to predict the evolution of the supermassive black holes that reside in
the galaxies in these regions. We find significant late time growth of black holes in massive galaxies,
although this growth may be exaggerated in our simulation. 
Our predicted $M_{\bullet}-M_{*, \mathrm{bulge}}$ relation agrees well with observed trends,
as might be expected since our accretion rate onto the black holes was set to reproduce
the $z=0$ SMBH mass densities in cluster and void regions determined from observations.
 We calculate the contribution to the mass
of each central black hole from accretion of gas and mergers with smaller black holes subsequent
to the mergers of their host galaxies. We find that in the cluster, the 
total BH mass from mergers is added later on average than the mass from accretion,
with half the mass from accretion added before $z \approx 1.0$ and half the
mass from mergers added after $z \approx 0.7$. In the void, half the total mass
accreted onto the BH population is accreted before $z \approx 0.6$, but half the merged
mass is added after $z \approx 0.5$. Mergers contribute
a negligible amount to the mass of black holes with 
$M_{\bullet} \lesssim 10^{6} M_{\odot}$ in both the cluster and
the void region. In the void, the number-weighted mean fraction of mass from mergers 
rises with black hole mass up to $\approx 5\%$ for
BHs with $M_{\bullet} \gtrsim 10^{8} M_{\odot}$. In the cluster,
the fraction from mergers reaches a maximum value of $23\%$ for 
$M_{\bullet} \approx 10^{10} M_{\odot}$.
The mean fraction of mass from mergers is larger in the cluster box than
for galaxies of the same mass in the void box, although values for
high-mass BHs have large scatter (Figure \ref{graph:graph2}).

Additionally, we predict the mass in black holes orbiting in galaxies
due to a galaxy-galaxy merger or gravitational wave recoil.
 While essentially negligible in the void, a significant amount of such mass is
expected in cluster galaxies with $M_{*, \mathrm{bulge}} \gtrsim 10^{11} M_{\odot}$ at $z = 0$. 
In the cluster, $40\%$ of the BHs by number and $7.6\%$ of the BH mass is orbiting.
We predict around 2 orbiting black holes on average for a galaxy in the cluster with bulge mass
around $10^{11} M_{\odot}$, or total stellar mass around $1.3 \times 10^{11} M_{\odot}$.
Such orbiting black holes are expected to produce observational signatures such as stellar tidal
disruption flares that are off-center in the galaxy \citep{komossa2008, li2012, liu2013}. 
They are also a candidate to explain via gaseous accretion 
observed ultra-luminous X-ray sources \citep{islam2004, volonteri2005b, mcwilliams2012}.

We compute the expected energy emitted in gravitational wave radiation due to black hole
mergers, shown in Figure \ref{graph:graph10}. More energy is expected at smaller redshifts due to the larger
amount of mass added via SMBH mergers during this time. We calculate the total strain amplitude from gravitational
waves to be $\log A = -14.7 \pm 0.1$. This is potentially observable by
pulsar timing arrays \citep{mcwilliams2012, sesana2013}. We also compute the bolometric luminosity from accretion,
shown in Figure \ref{graph:graph4}, which is directly proportional to the mass increase from 
accretion. 

We keep track of both SMBHs that end up orbiting in a galaxy due to insufficient dynamical friction
and SMBHs that are ejected from their hosts by gravitational wave recoils. These two populations
are not accounted for in attempts to calculate the local mass density of SMBHs by measuring the
masses of black holes at the centers of galaxies. As such, they are a correction to the Soltan argument.
We find that such SMBHs constitute between $6\%$ 
and $21\%$ of the total mass in SMBHs, with
a mean of $11.2\pm3.4\%$. 
Thus the efficiency of accretion in producing observable
energy output calculated by the Soltan argument should be reduced by the factor 
$0.888\pm0.034$.

We also find a decreasing variance around the $M_{\bullet}-M_{*, \mathrm{bulge}}$ relation
with increasing mass in both the cluster and the void, shown 
in Figure \ref{graph:graph3}. This is a result of
the fact that more massive SMBHs have undergone more accretion episodes, which tighten the scaling
relations due to the central limit theorem. Although current observational data are not sufficient to
confirm or disprove the existence of a decrease in scatter, our results are consistent
with the latest observational findings within the errors. 

Therefore, late time mergers and their environment have interesting and 
observationally detectable consequences for the mass assembly history 
of supermassive black holes. The prediction that we make of a significant
population of orbiting SMBHs in massive cluster galaxies is testable by future
observations.

\begin{acknowledgements}

Computing resources were in part provided by the NASA High-End
Computing (HEC) Program through the NASA Advanced
Supercomputing (NAS) Division at Ames Research Center.
This work is supported in part by grant NASA NNX11AI23G.
This work was also supported by World Premier International Research Center 
Initiative (WPI Initiative), MEXT, Japan.
A.K. is supported by the National Science Foundation Graduate Research
Fellowship, Grant No. DGE-1148900.
P.N. acknowledges support from a NASA-NSF TCAN award number 1332858.

\end{acknowledgements}

\bibliographystyle{apj}

\end{document}